\documentclass[11pt]{article}

\usepackage{jheppub}

\usepackage{graphicx}
\usepackage{epstopdf}
\usepackage{amsmath, amssymb}

\usepackage{bm}
\usepackage{bbold}
\usepackage{caption}
\usepackage{subcaption}

\newcommand{\be}{\begin{eqnarray}}
\newcommand{\ee}{\end{eqnarray}}

\newcommand{\nn}{\nonumber}
\newcommand{\bn}{\begin{enumerate}}
\newcommand{\en}{\end{enumerate}}

\def\IC{\mathbb{C}}

\def\IZ{\mathbb{Z}}

\def\CA{{\cal A}}

\def\CI{{\cal I}}

\def\CL{{\cal L}}
\def\CM{{\cal M}}
\def\CN{{\cal N}}

\def\CS{{\cal S}}
\def\CT{{\cal T}}

\def\CV{{\cal V}}

\def\a{\alpha}

\def\g{\gamma}

\def\half{\frac{1}{2}}

\def\p{\partial}

\newcommand{\ket}[1]{\left|{#1}\right\rangle}

\def\tr{{\rm Tr}}

\def\PE{{\rm PE}}

\def\vec#1{\bm{#1}}

\def\asu{\widehat{\mathfrak{su}}}
\def\su{\mathfrak{su}}

\title{Macdonald Index and Chiral Algebra}

\author{
Jaewon Song
}

\affiliation{Department of Physics, University of California, San Diego \\
 La Jolla, CA 92093, USA}

\emailAdd{jsong@physics.ucsd.edu}

\abstract{For any 4d $\CN=2$ SCFT, there is a subsector described by a 2d chiral algebra. The vacuum character of the chiral algebra reproduces the Schur index of the corresponding 4d theory. The Macdonald index counts the same set of operators as the Schur index, but the former has one more fugacity than the latter. We conjecture a prescription to obtain the Macdonald index from the chiral algebra. The vacuum module admits a filtration, from which we construct an associated graded vector space. From this grading, we conjecture a notion of refined character for the vacuum module of a chiral algebra, which reproduces the Macdonald index. We test this prescription for the Argyres-Douglas theories of type $(A_1, A_{2n})$ and $(A_1, D_{2n+1})$ where the chiral algebras are given by Virasoro and $\widehat{\mathfrak{su}}(2)$ affine Kac-Moody algebra. When the chiral algebra has more than one family of generators, our prescription requires a knowledge of the generators from the 4d.
}

\begin{document} 
\maketitle
\flushbottom

\section{Introduction}
As was shown in a beautiful work \cite{Beem:2013sza}, any four-dimensional $\CN=2$ superconformal field theory (SCFT) has a subsector described by a two-dimensional chiral algebra or vertex operator algebra. In this 4d/2d correspondence, a special set of $\frac{1}{4}$-BPS operators, called the Schur operators, are mapped to the states in the vertex operator algebra. Any flavor symmetry $\mathfrak{g}$ of 4d theory becomes an affine Kac-Moody symmetry $\hat{\mathfrak{g}}_{k_{2d}}$, with 
\begin{align}
 c_{2d} = -12 c_{4d} \ , \qquad k_{2d} = -\half k_{4d} \ .
\end{align}
Here $c_{4d}$ is one of the 4d conformal anomaly coefficients, $c_{2d}$ is the Virasoro central charge, $k_{4d}$ is the flavor central charge and $k_{2d}$ is the level of the affine Kac-Moody algebra. 
This correspondence has been further studied for the class $\CS$ theories \cite{Beem:2014rza,Lemos:2014lua}, Argyres-Douglas theories \cite{Cordova:2015nma, Buican:2015ina, Xie:2016evu}, $\CN=3$ SCFTs \cite{Nishinaka:2016hbw, Lemos:2016xke} and also in the context of gravity dual \cite{Bonetti:2016nma}. 
The chiral algebra structure also provides us predictions for generic $\CN=2$ SCFTs such as a bound on the central charge \cite{Liendo:2015ofa, Lemos:2015orc}, and on the dimension of the conformal manifold \cite{Buican:2016arp}. Similar chiral algebra structure was also found in 6d $\CN=(2, 0)$ SCFT \cite{Beem:2014kka}. 

The vacuum character of the chiral algebra reproduces the limit of the superconformal index \cite{Kinney:2005ej,Romelsberger:2005eg}, called the Schur index \cite{Gadde:2011ik,Gadde:2011uv}. The Schur index is defined as
\begin{align}
 \CI(q) = \tr (-1)^F q^{\Delta-R} \ ,
\end{align}
where $\Delta$ is the scaling dimension and $R$ is the Cartan of the $SU(2)_R$ symmetry. The trace is over the $\frac{1}{4}$-BPS states corresponding to the Schur operators.\footnote{Actually, the trace formula is still valid for the Schur index even if it is summed over all states. This is a special property of the Schur index not readily extends to other limits of the index.} 
Using the chiral algebra associated to the 4d $\CN=2$ SCFT, one can compute the Schur index for non-Lagrangian SCFTs such as Argyres-Douglas (AD) theories \cite{Argyres:1995jj,Argyres:1995xn}. AD theories tend to have a simple chiral algebra. For example, for the simplest AD theory called $H_0$ or $(A_1, A_2)$ theory, the corresponding chiral algebra is simply given by a Virasoro algebra with $c=-\frac{22}{5}$, which is the central charge for the Yang-Lee model, the simplest non-unitary minimal model. It turns out that the chiral algebra of the $(A_1, A_{2n})$ theory is given by that of the non-unitary Virasoro minimal model $\CM(2, 2n+3)$. For the $(A_1, D_{2n+1})$ theory, the chiral algebra is given by the $\asu(2)_{-\frac{4n}{2n+1}}$ affine Kac-Moody algebra.

The purpose of this paper is to extract more refined information on the spectrum of the 4d $\CN=2$ SCFT from the associated chiral algebra, namely the Macdonald index. The Macdonald index is defined as
\begin{align} \label{eq:MacIdxDef}
 \CI(q, t) = \tr_M (-1)^F q^{\Delta - 2R + r } t^{R-r} = \tr_M (-1)^F q^{\Delta-R} T^{R-r} \ , 
\end{align}
where $t=qT$ and the trace is over the states with $\Delta - 2j_1 - 2R - r = 0$ and $j_1 - j_2 +r = 0$.\footnote{Unlike the case of the Schur index, here we should take the trace only over the $\frac{1}{4}$-BPS states.} By setting $t=q$ or $T=1$, we recover the Schur index. Even though the Macdonald index contains more refined information than the Schur index with one more fugacity, it gets contributions from the same set of operators as the Schur index. Therefore it is plausible that one should be able to extract the Macdonald index from the chiral algebra. We claim that this is indeed possible at least when the chiral algebra is simple enough, such as the case of Virasoro or affine Kac-Moody algebra. We define a notion of the refined character for a representation of the chiral algebra that can be identified with the Macdonald index.

Suppose the chiral algebra $\CA$ has generators $ \{ X^{(i)}_{n} \}$ with $n \in \IZ$ labeling the modes and $(i)$ labeling different generators of the algebra. Then the vacuum module $\CV$ of the chiral algebra $\CA$ is constructed by applying strings of negative modes on the vacuum state: 
\begin{align}
 \CV = \textrm{span} \left\{ \left( \prod_{k} X^{(i_k)}_{-n_k} \right) \ket{0} \bigg{|} ~ n_{k} \in \IZ_{>0} \right \} / \{ \textrm{null states} \}\ , 
\end{align}
where the vacuum $\ket{0}$ is annihilated by the positive modes. The Verma module generated by the negative modes is generally reducible, due to the null states and their descendants. Therefore they have to be removed. The grading corresponds to $q$ in the Schur index or the character is given by the eigenvalue of the Virasoro generator $L_0$. Normally, this is obtained by taking the sum of the mode numbers, $\sum_k n_k$ for each state. 

Now, we conjecture that the other grading on the vacuum module counted by the fugacity $T$ is essentially given by the number of generators acted on the vacuum state. In other words, if we have a state given by $\prod_{k=1}^\ell X^{(i_k)}_{-n_k} \ket{0}$, it contributes $q^{\sum_{k} n_k} T^{\sum_k  w(X^{(i_k)}) }$ to the refined character, where $w(X^{(i)})$ depends on the type of the generator $X^{(i)}$ in $\CA$. We find that the generators for the current algebra $J^a_{n}$ has $w(J^a)=1$, and also the Virasoro generators $L_n$ has $w(L)=1$. 
For the general chiral algebra, the value of $w(X)$ is not uniquely determined and depends on the type of the generator $X$.\footnote{The author would like to thank Leonardo Rastelli for pointing this out.} Unless the chiral algebra is generated by one set of generators where one can simply fix $w(X)=1$, there is no natural choice in general. This makes our prescription incomplete. Once we know the 4d origin of the generators, it is possible to assign appropriate values for the $w(X)$, but we have not found an inherently 2d way of determining the weight.
For the most part of the current paper, we restrict ourselves to the case where $w(X)=1$. We find other values of $w(X)$, which will be discussed in section \ref{sec:Lag} and \ref{sec:conclusion}. 

This prescription is possible thanks to the filtration $\CV_0 \subset \CV_1 \subset \CV_2 \subset \ldots$ of the vacuum module $\CV$ of the chiral algebra, with 
\begin{align}
 \CV_k = \textrm{span} \{ X^{(i_1)}_{-n_1} X^{(i_2)}_{-n_2}  \ldots X^{(i_m)}_{-n_m} \ket{0} : n_1 \ge n_2 \ge \ldots \ge n_m,  m \le k \} /  \{ \textrm{null states} \}  \ . 
\end{align}
This filtration\footnote{A similar filtration and the notion of the `refined character' we define in the current paper has been already discussed in \cite{feigin2009pbw}. The author would like to thank the anonymous referee for pointing out this reference.} allows us to construct associated graded vector space $V_{gr}$
\begin{align}
 V_{gr} = \CV_0 \oplus \bigoplus_{i=1}^\infty (\CV_{i}/\CV_{i-1}) \ . 
\end{align}
Let us write $V^{(i)} \equiv \CV_i / \CV_{i-1}$ for $i\ge 1$ and $V^{(0)} \equiv \CV_0 = \{ \ket{0} \}$. 
Then we can decompose the $V_{gr}$ further in terms of the eigenspaces of $L_0$ as 
\begin{align}
 V_{gr} = \bigoplus_{i\ge 0} \bigoplus_{h} V^{(h)}_{i} \ , 
\end{align}
where $V^{(h)}_i$ is the eigenspace of $L_0$ with the eigenvalue $h$. 
The vector space $V^{(h)}_i$ do not form a $\CA$-module since any generators in $\CA$, but they do form $\mathfrak{g}$-modules under the finite symmetry algebra whenever there is one. There is one subtle thing regarding the filtration/grading. Even when the Virasoro generators are written in terms of other generators, we need to treat them also as one of the generators $X^{(i)}_{-n}$ (so that $X^{(i)}$ is either one of the generators of $\CA$ or $L$). The relation between generators (such as $L_n \sim \sum_m J_{n-m}J_m$) will render some linear combination of the vectors to be null, so we take them out as well. We will discuss this in greater detail in section \ref{sec:su2}.

Now, we define the refined character as
\begin{align}
 Z_{\CV}^{\textrm{ref}}(\vec{z}; q, T) = \sum_{i=0}^{\infty} \sum_{h} ch (V_{h}^{(i)}; \vec{z}) q^h T^i \ , 
\end{align}
where the $ch(V_{h}^{(i)}; \vec{z}) = \tr_{V_{h}^{(i)}}( \prod_a z_a^{F_a})$ denotes the character of the (non-affine) symmetry $\mathfrak{g}$ on $V_{h}^{(i)}$. Here $F_a$ are the Cartan generators of $\mathfrak{g}$.  
Our conjecture is that the Macdonald index for an $\CN=2$ SCFT $\CT$ is given by the refined character of the vacuum module $\CV_{\CT}$ of the associated chiral algebra $\CA[\CT]$
\begin{align} 
 \CI_{\CT}(q, t; z) \Big|_{t=qT} = Z_{\CV_{\CT}}^{\textrm{ref}} (\vec{z}; q, T) \ . 
\end{align}
Note that this extra grading has nothing to do with the global symmetry present in the 2d system. If we assign a particular charge to the generators, it is not going to give a consistent quantum number since the generators of $\CA$ satisfy non-linear algebraic relations. Moreover, we need to know explicit form of each null vectors to remove the contribution to the refined character in a correct way. We demonstrate this prescription of obtaining the refined character for some simple examples. Since the Argyres-Douglas theories have the simplest corresponding chiral algebra, we will use them as our main examples.  

Recent studies on AD theories allow us to test whether the refined character gives us the correct Macdonald index. A large class of generalized AD theories can be constructed from M5-branes wrapped on a Riemann surface \cite{Xie:2012hs,Xie:2013jc, Wang:2015mra}, many of which have rather simple chiral algebras \cite{Xie:2016evu}. From this M5-brane picture, Schur and Macdonald indices for the AD theories have been obtained from a topological field theory on a sphere with  irregular punctures \cite{Buican:2015ina,Buican:2015tda, Song:2015wta, Song:2017oew}. 
In addition, a relation between the BPS spectrum in the Coulomb branch and the Schur index has been discovered \cite{Cordova:2015nma,Cecotti:2015lab,Cordova:2016uwk}. Trace of the (inverse of) BPS monodromy operator \cite{Cecotti:2010fi,Iqbal:2012xm}, which is a wall-crossing invariant quantity, turns out to give the Schur index.  The full indices for the (subset) of AD theories have been obtained recently by using $\CN=1$ gauge theory \cite{Maruyoshi:2016tqk, Maruyoshi:2016aim, Agarwal:2016pjo} flowing to the fixed point described by the AD theory. We will use the simplest classes of the AD theories, namely $(A_1, A_{2n})$ and $(A_1, D_{2n+1})$ theories, where Macdonald indices are computed using two independent methods. Our result adds one additional method to compute the same quantity. 

The organization of this paper is as follows. In section \ref{sec:Virasoro}, we study the simplest chiral algebra, the Virasoro minimal model. The $(A_1, A_{2n})$ Argyres-Douglas theory is the one giving the desired chiral algebra. We explain the computation of the refined character in detail and compare against the Macdonald index computed using different methods. In section \ref{sec:su2}, we move on to the $\asu(2)$ affine Kac-Moody algebra, which can be obtained from $(A_1, D_{2n+1})$ theory. Here we discuss in detail how the relation between $\asu(2)$ generators and Virasoro generators affect the grading structure. In section \ref{sec:Lag}, we discuss Lagrangian theories. Lagrangian theories can be obtained by combining free vector multiplets and hypermultiplets upon gauging. We give a straight-forward interpretation of the Macdonald index in terms of the refined character. Finally, we conclude with some remarks and open questions. In the appendix, we summarize explicit expressions for the Macdonald indices of the relevant Argyres-Douglas theories.


\section{$(A_1, A_{2n})$ theory from Virasoro algebra} \label{sec:Virasoro}

In this section, we focus on the simplest example where the chiral algebra is given by that of the Virasoro minimal models. 

\paragraph{Null states of the vacuum module}
The chiral algebra for the $(A_1, A_{2n})$ theory is given by the Virasoro algebra with the central charge being identical to that of the non-unitary minimal model $\CM(2, 2n+3)$, given by
\begin{align}
 c = 1 - \frac{3(2n+1)^2}{2n+3} \ . 
\end{align}
For example, the central charges for $n=1, 2, 3$ are $c=-\frac{22}{5}, -\frac{68}{7}, -\frac{46}{3}$. 

Let us consider descendant states in a vacuum module of the minimal model $\CM(2, 2n+3)$. The states in the Verma module is constructed by acting the following operators on the highest-weight vector $\ket{h, c}$ with the highest weight $h$: 
\begin{align}
\begin{array}{ll}
 \textrm{level 0}\quad & 1 \\
 \textrm{level 1}\quad & L_{-1}  \\
 \textrm{level 2}\quad & L_{-1}^2, \quad L_{-2}  \\
 \textrm{level 3}\quad & L_{-1}^3 , \quad L_{-2} L_{-1} , \quad L_{-3} \\
 \textrm{level 4}\quad & L_{-1}^4 , \quad L_{-2}L_{-1}^2 , \quad L_{-3} L_{-1} , \quad L_{-2}^2 , \quad L_{-4}  \\
 \textrm{level 5}\quad & L_{-1}^5 ,  \quad L_{-2}L_{-1}^3, \quad L_{-3} L_{-1}^2 , \quad L_{-2}^2 L_{-1} , \quad L_{-3} L_{-2} , \quad L_{-4}L_{-1} ,\quad  L_{-5} 
\end{array}
\end{align}
For the vacuum module, $L_{-1} \ket{0, c}$ is a null state. Therefore any descendant states of $L_{-1} \ket{0, c}$ should be removed. Then we have the follows states (omitting $\ket{0, c}$):
\begin{align}
\begin{array}{ll}
 \textrm{level 0}\quad & 1 \\
 \textrm{level 1}\quad &   \\
 \textrm{level 2}\quad & L_{-2}  \\
 \textrm{level 3}\quad & L_{-3} \\
 \textrm{level 4}\quad & L_{-2}^2 , \quad L_{-4}  \\
 \textrm{level 5}\quad & L_{-3} L_{-2} , \quad L_{-5} \\
 \textrm{level 6}\quad & L_{-2}^3, \quad L_{-3}^2, \quad L_{-4}L_{-2} , \quad L_{-6} \\
 \textrm{level 7}\quad & L_{-3} L_{-2}^2 , \quad L_{-4} L_{-3}  , \quad L_{-5}L_{-2} , \quad L_{-7} \\
 \textrm{level 8}\quad & L_{-2}^4 , \quad L_{-3}^2 L_{-2} , \quad L_{-4}L_{-2}^2, \quad L_{-4}^2 , \quad L_{-5}L_{-3} ,\quad L_{-6}L_{-2},\quad L_{-8}
\end{array}
\end{align} 
At each level, there can be null states formed by taking certain linear combinations. One can show that there is a null state at level 4 and 5 for $c=-\frac{22}{5}$ 
\begin{align} \label{eq:VirNull4}
\left( - \frac{3}{5} L_{-4} + L_{-2}^2 \right) \ket{0, c=-\frac{22}{5}}  \ , \quad
 \left( - \frac{2}{5} L_{-5} + L_{-3} L_{-2} \right) \ket{0, c=-\frac{22}{5}}   \ . 
\end{align}
At level 6 and 7, there are 2 null states for $c=-\frac{22}{5}$, and one null state for $c=-\frac{68}{7}$. More explicitly, we have
\begin{align}
\begin{split}
& \left( - \frac{6}{5} L_{-6} - \frac{3}{5} L_{-4} L_{-2} + L_{-2}^3 \right) \ket{0, c=-\frac{22}{5}} \ , \\
& \left( - \frac{8}{5} L_{-6} + 2 L_{-4} L_{-2} + L_{-3}^2 \right) \ket{0, c=-\frac{22}{5}} \ , \\
& \left( - \frac{38}{49} L_{-6} - \frac{11}{7} L_{-4} L_{-2} -\frac{1}{7} L_{-3}^2 + L_{-2}^3 \right) \ket{0, c=-\frac{68}{7}} \ ,
\end{split}
\end{align}
and
\begin{align}
\begin{split}
& \left( - \frac{6}{5} L_{-7} + \frac{3}{5} L_{-5} L_{-2} + L_{-3}L_{-2}^2 \right) \ket{0, c=-\frac{22}{5}} \ , \\
& \left( -  L_{-7} +  L_{-5} L_{-2} + L_{-4}L_{-3} \right) \ket{0, c=-\frac{22}{5}} \ , \\
& \left( - \frac{19}{49} L_{-7} - \frac{4}{7} L_{-5} L_{-2} - \frac{5}{7} L_{-4} L_{-3} + L_{-3}L_{-2}^2 \right) \ket{0, c=-\frac{68}{7}} \ .
\end{split}
\end{align}
At level 8 and 9, one can show that there are 4 null states for $c=-\frac{22}{5}$ and 2 null states for $c=-\frac{68}{7}$ and 1 null state for $c=-\frac{46}{3}$. The null states for $c=-\frac{46}{3}$ is given as follows:
\begin{align}
& \left( -\frac{278}{81} L_{-8}+\frac{7}{9} L_{-4}^2 +\frac{4}{27} L_{-5} L_{-3}-\frac{88}{27} L_{-6}L_{-2}-\frac{4}{9} L_{-3}^2 L_{-2} -\frac{26}{9} L_{-4} L_{-2}^2 +L_{-2}^4 \right) \ket{0, -\frac{46}{3}} \nn \\
& \quad \bigg( -\frac{194}{81}  L_{-9}+\frac{14}{27} L_{-5}L_{-4}-\frac{2}{3} L_{-6}L_{-3}-\frac{35}{27} L_{-7}L_{-2}-\frac{1}{9} L_{-3}^3 \\
 & \qquad \qquad \qquad \qquad\qquad -\frac{17}{9} L_{-4} L_{-3}L_{-2}-\frac{2}{3} L_{-5}L_{-2}^2 +L_{-3} L_{-2}^3 \bigg) \ket{0, -\frac{46}{3}} \nn
\end{align}
 

\paragraph{Refined character}
Now, let us define the refined character for the Virasoro module $\CV$ by keeping track of the number of $L$ operators. Write
\begin{align}
 Z^{\textrm{ref}}_{\CV} (q, T) = \sum_{k \ge 0} N(c, k; T) q^k \ , 
\end{align}
where $k$ is the eigenvalue of $L_0$, and $n(c, k; T)$ is the `refined' number of states at level $k$ with extra grading
\begin{align}
 N(c, k; T) = \sum_{m = 1}^k n(c, k, m) T^m \ , 
\end{align}
where $n(c, k, m)$ is the number of descendant states constructed by acting $m$ of $L$'s. 

More formally, this can be understood as follows. Let us consider a filtration of the vector spaces 
\begin{align} \label{eq:VirFilt}
 \CV_0 \subset \CV_1 \subset \CV_2 \subset \CV_3 \subset \ldots  \ , 
\end{align} 
with 
\begin{align}
 \CV_n = \textrm{span} \{ ( L_{-i_1} L_{-i_2} \ldots L_{-i_m} \ket{0} : i_1 \ge i_2 \ldots \ge i_m, m \le n \} / \{ \textrm{null states} \} \ . 
\end{align}
From here, we can define the associated graded vector space $V_{gr}$ as
\begin{align} \label{eq:grV}
 V_{gr} = \bigoplus_{i \ge 0} V^{(i)} = \bigoplus_{i\ge0} \bigoplus_h V^{(i)}_h \ , 
\end{align}
where $V^{(i)} =  \CV_{i}/\CV_{i-1}$ and $\CV_{-1} = \{ \ket{0} \}$. By further decomposing $V^{(i)}$ in terms of $L_0$ eigenspace $V_h^{(i)}$ with eigenvalue $h$, we get the second identity. Now, the refined character is given by
\begin{align}
 Z^{\textrm{ref}}_{\CV}(q, T) = \sum_{i\ge 0} \sum_h \left( \textrm{dim} V^{(i)}_h \right) q^h T^i \ . 
\end{align}

Let us consider a simple example in the 2d context. For a Verma module of the Virasoro algebra (removing the $q^{h-c/24}$ piece), the refined character can be simply written as
\begin{align}
\begin{split}
 Z_{\textrm{Verma}} &= \frac{1}{(t; q)} = \frac{1}{(qT; q)} = \prod_{i=1}^\infty \frac{1}{1-q^i T} \\
  &= 1 + q T + q^2 (T^2 + T) + q^3 (T^3 + T^2 + T) + q^4(T^4 + T^3+2T^2 + T) + \cdots \ . 
\end{split}
\end{align}
This can be easily seen by giving fugacity $q^n T$ to $L_{-n}$ for each $n>0$. Combinatorially, we can write the character for a Verma module as 
\begin{align}
  Z^{\textrm{ref}}_{\textrm{Verma}} (q) = \sum_{n \in \IZ_{\ge 0}} p(n) q^n\ , 
\end{align}
where $p(n)$ is the number of the integer partitions of $n$. Each partition can be represented by a Young diagram. We can interpret the fugacity $T$ as the fugacity counting the heights of the Young diagrams. So we write 
\begin{align}
 Z^{\textrm{ref}}_{\textrm{Verma}} (q, T) = \sum_{n \in \IZ_{\ge 0}} \sum_{k \le n} p(n, k) q^n T^k  \ , 
\end{align}
where $p(n, k)$ denotes the number of Young diagrams of $n$ boxes of height $k$.

In order to compute the refined character, we need to take the null states into account. The prescription we give is: whenever there is a null-state providing a relation between the states with a different number of $L$'s, we drop the state with the largest number of $L$'s. This prescription is compatible with the filtration \eqref{eq:VirFilt} because null state condition allows us to write an element of the $\CV_{i}$ in terms of the elements in $\CV_{j<i}$ that have a smaller number of $L$'s. For example, when $c=-\frac{22}{5}$, we have a null state at level $4$ as \eqref{eq:VirNull4}. This relation allows us to write $L_{-2}^2 \ket {0} \in \CV_2$ in terms of $L_{-4}\ket{0} = \frac{5}{3} L_{-2}^2 \ket{0} \in \CV_1 \subset \CV_2$. Therefore, in the associated graded vector space $V_{gr}$, the equivalence class of the $L_{-4}\ket{0}$ state exists in $V_1$, but that of $L_{-2}^2 \ket{0}$ is absent in $V_2 \subset V$. 

Using our prescription, we obtain
\begin{align}
\begin{array}{lll}
 N(-\frac{22}{5}, 4; T) = T, & N(-\frac{22}{5}, 5; T) = T, & N(-\frac{22}{5}, 6; T) = T + T^2 , \\
 N(-\frac{68}{7}, 4, T) = T+T^2 , & N(-\frac{68}{7}, 5; T) = T+T^2 , & N(-\frac{68}{7}, 6; T) = T + 2T^2 \ .
\end{array}
\end{align}
We find the refined vacuum characters for $c=-\frac{22}{5}, -\frac{68}{7}, -\frac{46}{3}$ to be
\begin{align}
 Z_{\CV(-\frac{22}{5})} &= 1 + q^2 T + q^3 T + q^4 T + q^5 T + q^6 (T + T^2) + q^7 (T + T^2) + q^8 (2T^2 + T) + O(q^9) , \nn \\
 Z_{\CV(-\frac{68}{7})} &= 1 + q^2 T + q^3 T + q^4 (T+T^2) + q^5 (T+T^2) + q^6 (T+2T^2) + q^7 (T + 2 T^2) + O(q^8) , \nn \\
 Z_{\CV(-\frac{46}{3})} &= 1 + q^2 T + q^3 T + q^4 (T+T^2) + q^5 (T+T^2) + q^6 (T+2T^2+ T^3)  \\ 
  & \qquad \qquad + q^7 (T + 2 T^2 + T^3) + O(q^8) \ , 
\end{align}
where $\CV(c)$ is the vacuum module for the Virasoro algebra with central charge $c$. 
They agree with the Macdonald indices \eqref{eq:MacIdx} of $(A_1, A_2)$, $(A_1, A_4)$ and $(A_1, A_6)$ theories upon identifying $t=qT$. 

Even though it is not computationally too difficult to compute the refined character to high orders in $q$, it would be desirable to find a closed form expression for the minimal models and prove that it agrees with the expression for the Macdonald index \eqref{eq:MacIdx}.

\section{$(A_1, D_{2n+1})$ theory from $\widehat{\mathfrak{su}}(2)$ affine Kac-Moody algebra} \label{sec:su2}
The chiral algebra associated to the $(A_1, D_{2n+1})$ Argyres-Douglas theory is conjectured to be given by the affine Kac-Moody algebra (or affine vertex operator algebra) $\widehat{\mathfrak{su}}(2)_k$ of level $k=-\frac{4n}{2n+1}$. 
Similar to the case of the Virasoro algebra, we conjecture that the fugacity $T$ counts the number of $L$ and $J$'s used to construct the states in the $\widehat{\mathfrak{su}}(2)$ module. But here we have additional subtlety. As we will see, this is because the Virasoro stress-energy tensor (or the conformal vector in the affine vertex operator algebra) is given in terms of the affine current. 

The commutation relations for the generators of $\widehat{\mathfrak{su}}(2)_k$ algebra are given as
\begin{align} 
\begin{split}
& [J_m^0, J_n^0 ] = \frac{k}{2} m\delta_{m+n, 0} ,  \\
& [J_m^0, J_n^\pm] = \pm J^{\pm}_{m+n} , \\
& [J_m^+, J_n^-] = 2 J^0_{m+n} + k m \delta_{m+n, 0} . 
\end{split} 
\end{align}
The vacuum state satisfies
\begin{align}
 J^+_m \ket{0, k} &= 0 \quad (m\ge 0) , \\
 J^0_m \ket{0, k} &= 0 \quad (m \ge 0)  , \\
 J^-_m \ket{0, k} &=0 \quad (m > 0) . 
\end{align}
The vacuum module is constructed by acting $J_{m}^{\pm, 0}$ operators with $m<0$ on the vacuum state $\ket{0, k}$. Note that even though $J^-_0$ does not annihilate the vacuum state, $J^-_0 \ket{0, k}$ has zero norm so that we remove this state and its descendants. 

For a string of $J$ operators acting on the vacuum state, we count the number of $J$ to give $T^n$ term in the refined index. In addition, we need to take care of the Virasoro generators $L_n$. For the Argyres-Douglas theories of type $(A_1, D_{2n+1})$, the stress-energy tensor of the chiral algebra is given in terms of the $\widehat{\mathfrak{su}}(2)_k$ generators $J_n^0, J_n^\pm$ via Sugawara construction
\begin{align}
 L_n = \frac{1}{2(k+2)} \sum_{m \in \IZ} \left( : J^0_m J^0_{n-m} + \half \left( J^+_m J^-_{n-m}  + J^-_m J^+_{n-m} \right) :\right) \ ,
\end{align}
where $: AB:$ stands for the usual normal ordering which puts the lower modes in front. The counting becomes more involved since $L \sim J^2$ so that we may overcount the number of $T$. Therefore we impose an additional rule for counting $T$; whenever there is a state that can be written in terms of $L$'s instead of $J^2$, we count it as $T^1$, not as $T^2$. 

As we have discussed in the introduction, this prescription can be made more precise by considering a filtration $\CV_0 \subset \CV_1 \subset \CV_2 \subset \ldots$, where 
\begin{align}
 \CV_k = \textrm{span} \{ X^{(i_1)}_{-n_1} X^{(i_2)}_{-n_2}  \ldots X^{(i_m)}_{-n_m} \ket{0} : n_1 \ge  \ldots \ge n_m, m \le k \} /  \{ \textrm{null states \& relations} \} . 
\end{align}
Here $X^{(i)}$ is one of the $\widehat{\mathfrak{su}}(2)$ generators $J^+, J^-, J^0$ or the Virasoro generator $L$. We remove the null states and impose the relations between Virasoro and affine current generators as well. Now we can construct associated graded vector space $V_{gr} = \bigoplus_i V^{(i)} = \bigoplus_i (\CV_{i}/\CV_{i-1} ) $ as \eqref{eq:grV} to define the refined character as
\begin{align}
 Z^{\textrm{ref}}_{\CV}(z; q, T) = \sum_{i\ge 0} \sum_h  ch(V^{(i)}_h;  z) q^h T^i \ . 
\end{align}
The only difference here from the Virasoro case is that now $V^{(i)}_h$ itself is a module of the finite $\mathfrak{su}(2)$ algebra so that we replace the dimension to the character $ch(V_h^{(i)}; z) = \tr_{V_h^{(i)}} z^{J^0_0}$. 

Now, let us perform this computation explicitly for the first few levels. We find that our prescription correctly reproduces the Macdonald index for the AD theory of type $(A_1, D_{2n+1})$. 

\paragraph{Level 1}
At level 1, the descendant states are given by
\begin{align}
 J^{\pm, 0}_{-1} \ket{0, k} \ . 
\end{align}
From counting $J$'s, we find the refined character at level 1 is given by 
\begin{align}
 Z^{(1)} (z; q, T) = q T \left( z^2 + \frac{1}{z^2} + 1 \right) = q T \chi_3 (z) ,
\end{align}
where $\chi_{2j+1}$ denotes the character of the spin-$j$ representation $R_j$ of the finite Lie algebra $\su(2)$ given by 
\begin{align}
 \chi_{2j+1}(z) = \tr_{R_j} z^{J^0} = \frac{z^{2j+1} - z^{-2j-1}}{z-z^{-1}} \ . 
\end{align}
 We see that the refined character at level 1 indeed agrees with the Macdonald index for the $(A_1, D_{2n+1})$ theory at order $q^1$.  At this level, Sugawara relation does not play a role in counting $T$. This is because 
\begin{align}
L_{-1} \sim \sum_a J_{-1}^a J_0^a + J_{-2}^a J_1^a + \cdots \ ,
\end{align}
so that $L_{-1}$ annihilates the vacuum state. 

\paragraph{Level 2}
At level 2, we have the following descendant states (excluding the descendants of the null state $J^-_0 \ket{0, k}$):
\begin{align}
\begin{array}{c|c}
\textrm{descendant state} & \textrm{refined character} \\
\hline \hline
 J_{-1}^+ J_{-1}^+, J_{-1}^+ J_{-1}^0, J_{-1}^0 J_{-1}^0, J_{-1}^+J_{-1}^- , J_{-1}^- J_{-1}^0, J_{-1}^- J_{-1}^- & q^2 T^2 \left(\chi_5 + \chi_1  \right) \\
 J_{-2}^+, J_{-2}^0, J_{-2}^- & q^2 T \chi_3 \\
  \hline
 L_{-2} \sim  J_{-1}^0 J_{-1}^0 + J_{-1}^+J_{-1}^- & q^2 (T-T^2)  
\end{array}
\end{align}
The operators on the left column act on the vacuum state. Here we omitted writing $\ket{0, k}$ on the left-hand side of the table. The refined character basically counts the number of $J$ operators, but with one significant modification due to the relation between the generators of the chiral algebra. 
The last row comes from the Sugawara relation 
\begin{align} \label{eq:su2vir2}
 L_{-2} \sim J_{-1}^0 J_{-1}^0 + J_{-1}^+ J_{-1}^- + \ldots \ , 
\end{align}
where we omitted the terms with positive modes that annihilate the vacuum state. From this relation, we get $T^1$ for the state $L_{-2} \ket{0} \sim (J_{-1}^0 J_{-1}^0 + J_{-1}^+ J_{-1}^-)\ket{0} \in \CV_1$ and $T^2$ for the orthogonal direction $(J_{-1}^0 J_{-1}^0 - \a J_{-1}^+ J_{-1}^-) \in \CV_2$ for some $\a$. This has an effect of subtracting $q^2 T^2$ from the refined character and adding $q^2 T^1$ instead. Notice that this relation does not affect the ordinary character given by the $T \to 1$ limit. 

We compute the Kac matrix at level 2, and we find that there is no null state except when $k=0, -2$. Therefore the refined character is given by
\begin{align}
 Z^{(2)} (z; q, T) = q^2 \left( T ( \chi_1  + \chi_3) + T^2 \chi_5 \right) \qquad (k \neq 0, -2) \ , 
\end{align}
which agrees with the Macdonald index of the $(A_1, D_{2n+1})$ theory. 

\paragraph{Level 3}
At level 3, we get the states as in table \ref{table:su2k3}. 
\begin{table} 
\begin{align}
\begin{array}{c|c}
\textrm{descendant states} & \textrm{refined character} \\
\hline \hline
	J_{-1}^+ J_{-1}^+ J_{-1}^+ & \\
	J_{-1}^+ J_{-1}^+ J_{-1}^0  &  \\
	 J_{-1}^+ J_{-1}^0 J_{-1}^0, J_{-1}^+ J_{-1}^+ J_{-1}^- & \\
	J_{-1}^0 J_{-1}^0 J_{-1}^0 , J_{-1}^0 J_{-1}^+ J_{-1}^- & q^3 T^3 (\chi_7 + \chi_3)  \\
	J_{-1}^- J_{-1}^0 J_{-1}^0, J_{-1}^- J_{-1}^+ J_{-1}^-  &  \\
	J_{-1}^- J_{-1}^- J_{-1}^0 & \\
	J_{-1}^- J_{-1}^- J_{-1}^- & \\
	\hline
	J_{-2}^+ J_{-1}^+ & \\
	J_{-2}^+ J_{-1}^0, J_{-2}^0 J_{-1}^+ &\\
	J_{-2}^0 J_{-1}^0, J_{-2}^+ J_{-1}^-, J_{-2}^- J_{-1}^+ & q^3 T^2( \chi_5 + \chi_3 + \chi_1) \\
	J_{-2}^- J_{-1}^0, J_{-2}^0 J_{-1}^-  &  \\
	J_{-2}^- J_{-1}^- & \\
	\hline
	J_{-3}^{\pm, 0} & q^3 T \chi_3 \\
	\hline \hline
	L_{-2} J_{-1}^{\pm, 0}  \sim (J_{-1}^0 J_{-1}^0 + J_{-1}^+ J_{-1}^-)J_{-1}^{\pm, 0}  & q^3 (T^2-T^3) \chi_3 \\
	L_{-3} \sim J_{-2}^0 J_{-1}^0 + \half J_{-2}^+ J_{-1}^- + \half J_{-2}^- J_{-1}^+ & q^3 (T-T^2) \chi_1
\end{array} \nn
\end{align}
\caption{The descendant states of a generic vacuum module at level 3 and their contribution to the refined character. The last two rows account for the relation among the Virasoro and affine current generators. We subtract the states that have the higher powers in $T$ and then replace it by the one that has lower powers in $T$.}
\label{table:su2k3}
\end{table}
At this level, we also have a relation among the generators 
\begin{align}
 L_{-3} \sim J_{-2}^0 J_{-1}^0 + \half J_{-2}^+ J_{-1}^- + \half J_{-2}^- J_{-1}^+ + \ldots \ . 
\end{align} 
Therefore, we have to assign $T^1$ instead of $T^2$ for the state $(J_{-2}^0 J_{-1}^0 + \half J_{-2}^+ J_{-1}^- + \half J_{-2}^- J_{-1}^+ )\ket{0} \sim L_{-3} \ket{0} $. 

Whenever there is a null vector, we have to mod it out from the module. Up to level 2, there is no null vector for the negative levels we are interested. 
At level 3, there is a null vector that has $\mathfrak{su}(2)$ weight $0$ when level $k$ is given by $k=0, 1, 2$ or $k=-2, -\frac{4}{3}$. When $k=-\frac{4}{3}$, we get the following null vector: (See also \cite{Gaberdiel:2001ny})
\begin{align} \label{eq:level3null0}
 \left( J^0_{-3} + \frac{3}{2} J^+_{-2} J^-_{-1} - \frac{3}{2} J^+_{-1} J^-_{-2} - \frac{9}{2} J^0_{-2} J^0_{-1} + \frac{9}{2} J^0_{-1} J^+_{-1} J^-_{-1} + \frac{9}{2} J^0_{-1} J^0_{-1}J^0_{-1} \right) \ket{0,  k=-\frac{4}{3}} 
\end{align}
There is also a null vector of $\su(2)$ weight $1$ (highest-weight state of spin-$1$ representation) for $k=0, 1, 2$ and $k=-2, -\frac{4}{3}$. When $k=-\frac{4}{3}$, we get
\begin{align} \label{eq:level3null1}
\left( \frac{5}{9} J^+_{-3} - \frac{2}{3} J^+_{-2} J^0_{-1} - \frac{1}{3} J^0_{-2} J^+_{-1} + J^+_{-1} J^0_{-1} J^0_{-1} + J^+_{-1} J^+_{-1} J^-_{-1} \right) \ket{0, k=-\frac{4}{3}}  \ . 
\end{align}
By complex conjugation, there is also a null state with $\su(2)$ weight $-1$. 
There is no null state with $\su(2)$ weight $2$ for the level other than $k=1, 2$. Therefore we have a triplet of states that become null for $k=-4/3$. 

The triplet of null states above will remove $T^2 \chi_3$ from the refined character. We remove $T^2$ instead of $T^3$ since the last two terms in \eqref{eq:level3null0} and \eqref{eq:level3null1} can be written as $\frac{9}{2} J_{-1}^0 L_{-2}$ and $J_{-1}^+ L_{-2}$ respectively.  
Summarizing, we obtain the refined character at level 3 as 
\begin{align}
 Z^{(3)} (z; q, T) &= 
 \begin{cases} q^3 \left( T ( \chi_1  + \chi_3) + T^2 (\chi_3 + \chi_5 ) +  T^3 \chi_7 \right) & (k=-4/3)\ , \\ 
q^3 \left( T ( \chi_1 + \chi_3) + T^2 ( 2 \chi_3 + \chi_5 ) +  T^3 \chi_7 \right)  & (k \neq 0, 1, 2, -2, -4/3)  \ . 
\end{cases}
\end{align}
This result agrees with the Macdonald index for the $(A_1, D_3)$ and other $(A_1, D_{2n+1})$ theories.  

\paragraph{Level 4}
At level 4, we have the states as given in table \ref{table:su2level4}.
\begin{table}
\begin{align}
\begin{array}{c|c||c}
\textrm{mode numbers} & \textrm{type of generators $J^{\pm, 0}$} & \textrm{refined character} \\
\hline \hline
 J_{-4} & (+), (0), (-) & T \chi_3 \\
 \hline
 J_{-3} J_{-1} & (++), (+0), (00), (-0), (--) & T^2 \chi_5 \\
   & (0+), (+-), (-+), (0-) & T^2 \chi_3 + T^2 \chi_1 \\
 \hline
 (J_{-2})^2  & (++), (+0), (00), (-0), (--) & T^2 \chi_5 \\
  & (+-) & T^2 \chi_1 \\
 \hline
 J_{-2} (J_{-1})^2 & (+++), (++0), (+00), (000), (-00), (--0), (---) & T^3 \chi_7 \\
  & (0++), (00+), (+-0), (00-), (0--) & T^3 \chi_5  \\
  & (+-+), (-++), (-+0), (0+-), (+--), (-+-) & 2 T^3 \chi_3 \\
 \hline
 (J_{-1})^4 & (+^4), (+^3 0) (+^2 0^2) (+0^3) (0^4) (-0^3)(-^2 0^2)(-^3 0)(-^4) & T^4 \chi_9  \\
 & (+++-), (+0+-),  (00+-), (-0+-), (--+-) & T^4 \chi_5 \\
 & (+-+-) & T^4 \chi_1 \\
 \hline \hline
 L_{-4} & & (T-T^2) \chi_1 \\
 L_{-3} J_{-1} & (+), (0), (-) & (T^2-T^3) \chi_3 \\
 L_{-2} J_{-2} & (+), (0), (-) & (T^2-T^3) \chi_3 \\
 \hline
 L_{-2} (J_{-1})^2 & (++), (+0), (00), (+-), (-0), (--) & (T^3-T^4) (\chi_5+\chi_1) \\
 L_{-2} L_{-2} & & (T^2-T^3) \chi_1 
\end{array} \nn
\end{align}
\caption{States of a generic vacuum module at level 4. The first two columns of the table encode a sequence of generators being acted on the vacuum state. The first column denotes the sequence of generators with the mode number given via subscript. Each item in the second column encodes the type of $J \in \{J^+, J^-, J^0 \} $. For example, $J_{-2} (J_{-1})^2$ with $(+-0)$ refers to the state $J_{-2}^+ J_{-1}^- J_{-1}^0 \ket{0}$. The superscripts in the second column are used to denote multiplicity. For example, $(J_{-1})^4$ with $(+^4)$ refers to $(J_{-1}^+)^4 \ket{0}$.
The last five rows take care of the Sugawara relation by subtracting the states that are overcounted. } 
\label{table:su2level4}
\end{table}
At level 4, we have a Sugawara relation
\begin{align}
 L_{-4} \sim J_{-3}^0 J_{-1}^0 + \half J_{-3}^+ J_{-1}^- + \half J_{-3}^- J_{-1}^+ +J_{-2}^0 J_{-2}^0  + J_{-2}^+ J_{-2}^- +  \ldots \ . 
\end{align}
Some states can be generated by $L_{-2, -3, -4}$ which replaces the naive weight for $T$ that is given by simply counting the number of $J$'s. See the last five rows in table \ref{table:su2level4}. Here, notice that the last entry $L_{-2}L_{-2} \ket{0}$ in the table contributes to the refined character by $T^2-T^3$. This is the case because there is a vector given by a linear combination of $L_{-2} (J_{-1})^2 \ket{0} \in \CV_3$ which is identical to $L_{-2}L_{-2} \ket{0} \in \CV_2$ via Sugawara relation. We should be careful not to over-subtract the states that have been already absorbed into the vector of the form $L_{-2} (J_{-1})^2 \ket{0}$. Here we are identifying the state of the form $(J_{-1}^0 J_{-1}^0 + J_{-1}^+ J_{-1}^-)^2 \ket{0} \in \CV_4$ with the state $(J_{-1}^0 J_{-1}^0 + J_{-1}^+ J_{-1}^-) L_{-2} \ket{0} \in \CV_3$ and then finally with the state $L_{-2}L_{-2} \ket{0} \in \CV_2$.

When there is no null state, we obtain the refined character at level 4 by adding the terms in the third column of table \ref{table:su2level4}. 
There are null states at level 4 with $\su(2)$ weight $2$ for $k=0^{(6)}, 1^{(5)}, 2^{(3)}, 3^{(1)}$ (the superscript inside the parenthesis denotes the multiplicity of a null vector) and $k=-2, -\frac{1}{2}, -\frac{4}{3}$. There are also null vectors with $\su(2)$ weight $1, 0, -1, -2$. When $k=-\frac{4}{3}$, we get a null state
\begin{align}
\left( \frac{8}{9} J^+_{-3} J^+_{-1} - \frac{1}{3} J^0_{-2}  J^+_{-1} J^+_{-1} - 
 \frac{2}{3} J^+_{-2} J^+_{-1} J^0_{-1} + 
 (J^+_{-1})^2 (J^0_{-1})^2 + 
 (J^+_{-1})^3 J^-_{-1} \right) \ket{0, k=-\frac{4}{3}} , 
\end{align}
which will remove the $q^4 T^3 \chi_5$ from the refined index. It removes $T^3$ instead of $T^4$ because the last two terms can be written as $(J_{-1}^+)^2 L_{-2}$. 

There are also null states with weight $1$ for $k=0^{(10)}, 1^{(7)}, 2^{(3)}, 3$ and $k=-2^{(3)}, -\frac{1}{2}, (-\frac{4}{3})^{(2)}$. When $k=-\frac{4}{3}$ we have two null states given by $\vec{x}_{\textrm{null}} \cdot \vec{v}_1$ with
\begin{align}
\vec{x}_{\textrm{null}} \in \left\{ \left( \frac{2}{9},-\frac{1}{9},0,\frac{5}{3},\frac{1}{3},-\frac{1}{3},1,0,1,1\right) ,\left(-\frac{2}{3},-\frac{10}{3},\frac{10}{3},3,1,2,2,1,0,0\right) \right\} \ , 
\end{align}
and
\begin{align}
\begin{split}
\vec{v}_1 =& \Big( 
J^+_{-4},~J^+_{-3}J^0_{-1},~J^0_{-3}J^+_{-1},~J^+_{-2}J^0_{-2},~J^+_{-2}(J^0_{-1})^2,~J^0_{-2}J^+_{-1}J^0_{-1}, \\
 & { } \qquad J^+_{-2}J^-_{-1}J^+_{-1},~J^-_{-2}J^+_{-1}J^+_{-1},~J^+_{-1}(J^0_{-1})^3, ~J^+_{-1}J^0_{-1}J^+_{-1}J^-_{-1}
\Big) \ket{0, k=-\frac{4}{3}} \ . 
\end{split} 
\end{align}
The first null vector is coming from the descendant state of weight $2$, and the second null vector remove $q^4 T^2 \chi_3$. It gets $T^2$ instead of $T^3$ since
\begin{align}
\begin{split}
 & \left( J_{-2}^+ J_{-1}^0 J_{-1}^0 +2 J_{-2}^0 J_{-1}^+ J_{-1}^0 + 2 J_{-2}^+ J_{-1}^- J_{-1}^+ +J_{-2}^+ J_{-1}^- J_{-1}^+ + J_{-2}^- J_{-1}^+ J_{-1}^+ \right) \ket{0, k=-\frac{4}{3}} \\
 & \qquad \sim \left( J_{-2}^+  L_{-2} + 2 L_{-3} J_{-1}^+ \right) \ket{0, k=-\frac{4}{3}}  \ . 
\end{split}
\end{align}

Null states with weight 0 exist for $k=-\frac{1}{2}, -\frac{4}{3}$. There are 3 null states for $k=-\frac{4}{3}$, which are given by $\vec{x}_{\textrm{null}} \cdot \vec{v}_0$ with
\begin{align} \begin{split}
\vec{x}_{\textrm{null}} =& \left(-10,\frac{52}{9},\frac{28}{9},-\frac{50}{9},-\frac{5}{3},\frac{8}{3},6,-\frac{2}{3},\frac{14}{3},0,-1,0,1\right) , \\ 
& \left(\frac{2}{3},-\frac{4}{9},\frac{1}{3},\frac{1}{3},0,-\frac{2}{3},-1,\frac{1}{3},-\frac{1}{3},0,1,1,0\right), \\
& \left(-\frac{8}{3},0,\frac{5}{3},-\frac{5}{3},-1,0,3,1,1,1,0,0,0\right)
\end{split} \end{align}
and 
\begin{align}
\begin{split}
\vec{v}_0 =& \Big( J^0_{-4}, ~J^0_{-3}J^0_{-1}, ~J^+_{-3}J^-_{-1}, ~J^-_{-3}J^+_{-1}, ~J^0_{-2} J^0_{-2}, ~J^+_{-2}J^-_{-2}, ~J^0_{-2}(J^0_{-1})^2, ~J^+_{-2}J^-_{-1}J^0_{-1}, \\
&~ J^-_{-2}J^+_{-1}J^0_{-1},~J^0_{-2}J^+_{-1}J^-_{-1},~ (J^0_{-1})^4, ~(J^0_{-1})^2 J^+_{-1}J^-_{-1},~J^+_{-1}J^-_{-1}J^+_{-1}J^-_{-1} \Big) \ket{0, k=-\frac{4}{3}} .
\end{split}
\end{align}
The first and the second null vector comes from the descendant states of weight $2$ and $1$. The third null vector removes $q^4 T^2 \chi_1$. We get $T^2$ instead of $T^3$ since
\begin{align}
 J_{-2}^0 J_{-1}^+ J_{-1}^- +  J_{-2}^- J_{-1}^+ J_{-1}^0 + J_{-2}^+ J_{-1}^- J_{-1}^0 + 3  J_{-2}^0 J_{-1}^0 J_{-1}^0
  \sim  J_{-2}^0 L_{-2} + 2 L_{-3} J_{-1}^0 \ . 
\end{align}

Summarizing, we get the refined character at level 4 as
\begin{align}
 Z^{(4)} =
 \begin{cases}
  q^4 \left( T ( \chi_1 + \chi_3) + T^2 (  \chi_1 + 2 \chi_3 + 2 \chi_5 ) +  T^3 (\chi_5 + \chi_7 ) + T^4 \chi_9 \right) & (k=-\frac{4}{3}), \\ 
  q^4 \left( T ( \chi_1 + \chi_3) + T^2 ( 2 \chi_1 + 3 \chi_3 + 2 \chi_5 ) +  T^3 (2 \chi_5 + \chi_7 ) + T^4 \chi_9 \right)  & (k \notin N_4), 
  \end{cases}
\end{align}
where $N_4 = \{0, 1, 2, 3, -2, -\frac{1}{2}, -\frac{4}{3} \}$. It agrees with the Macdonald index for the $(A_1, D_{2n+1})$ theory. 

\paragraph{Level 5}
At level 5, we have the states as given in table \ref{table:su2level5}. 
\begin{table}
\begin{align}
\begin{array}{c|c||c}
\textrm{modes} & \textrm{type of generators } J^{\pm, 0} & \textrm{refined character} \\
\hline \hline
 J_{-5} & (+), (0), (-) & T \chi_3 \\
 \hline
 J_{-4} J_{-1} & (++), (+0), (00), (-0), (--) & T^2 \chi_5 \\
   & (0+), (+-), (-+), (0-) & T^2 \chi_3 + T^2 \chi_1 \\
 \hline
 J_{-3} J_{-2} & (++), (+0), (00), (-0), (--) & T^2 \chi_5 \\
   & (0+), (+-), (-+), (0-) & T^2 \chi_3 + T^2 \chi_1 \\
 \hline
 J_{-3} (J_{-1})^2 & (+++), (++0), (+00), (000), (-00), (--0), (---) & T^3 \chi_7 \\
  & (0++), (00+), (+-0), (00-), (0--) & T^3 \chi_5  \\
  & (+-+), (-++), (-+0), (0+-), (+--), (-+-) & 2T^3 \chi_3 \\
 \hline
 (J_{-2})^2 J_{-1} & (+++), (++0), (+00), (000), (-00), (--0), (---) & T^3 \chi_7 \\
  & (+0+), (00+), (0+-), (00-), (0--) & T^3 \chi_5  \\
  & (+-+), (++-), (0-+), (+-0), (-+-), (--+) & 2 T^3 \chi_3 \\
 \hline
 J_{-2} (J_{-1})^3 & (+^4), (+^3 0) (+^2 0^2) (+0^3) (0^4) (-0^3)(-^2 0^2)(-^3 0)(-^4) & T^4 \chi_9  \\
 & (0+^3), (0++0), (0+00), (+-00), (0-00), (0--0), (0-^3) & T^4 \chi_7 \\
 & (+++-) , (0++-),   (00+-), (0-+-), (--+-) & T^4 \chi_5 \\
 & (-+++), (-++0), (-+00),  (-+-0), (-+--) & T^4 \chi_5 \\
 & (+-+0), (-++-), (-+--), & T^4 \chi_3  \\
 & (+-+-) & T^4 \chi_1 \\
 \hline
 (J_{-1})^5 & (+^5)(+^4 0)(+^3 0^2)(+^2 0^3) (+0^4)(0^5)(-0^4)(-^2 0^3)(-^3 0^2)(-^4 0)(-^5) & T^5 \chi_{11} \\
 & (+^4-)(+^3 - 0)(+^2 -0^2)(+-0^3)(+-^2 0^2)(+-^30)(+-^4) & T^5 \chi_7 \\
 & (+-+-+)(+-+-0) (+-+--)& T^5 \chi_3 \\
 \hline \hline
 L_{-5} & & (T-T^2)\chi_1 \\
 L_{-4} J_{-1}  & (+), (0), (-) & (T^2 - T^3) \chi_3  \\
 L_{-3} J_{-2} & (+), (0), (-) & (T^2 - T^3) \chi_3 \\
 L_{-2} J_{-3} &(+), (0), (-) & (T^2 - T^3) \chi_3 \\
 \hline
 L_{-3} (J_{-1})^2 & (++), (+0), (00), (+-), (-0), (--) & (T^3 - T^4) (\chi_5 + \chi_1)\\
  L_{-3} L_{-2}  & & (T^2 - T^3) \chi_1 \\
  \hline
  L_{-2} (J_{-1})^3 & (+^3)(+^20)(+0^2)(+^2-)(0^3)(+-0)(-0^2)(-^2+)(-^20)(-^3) & (T^4 - T^5)(\chi_7 + \chi_3) \\
    L_{-2} L_{-2} J_{-1} & (+), (0), (-) & (T^3 - T^4) \chi_3 \\
 \hline
 L_{-2} J_{-2} J_{-1} & (++), (+0), (0+), (00), (+-), (-+), (-0), (0-), (--) & (T^3 - T^4) (\chi_5 + \chi_3+\chi_1) \\
  L_{-3} L_{-2} ~(*) & & - (T^3 - T^4) \chi_1 
\end{array} \nn 
\end{align} \caption{States of a generic vacuum module at level 5. One should be careful here not to overcount the states. The last entry with (*) is there to remove one superfluous state that is given by a linear combination $L_{-2} J_{-2}J_{-1}\ket{0}$, which is already counted as $L_{-3}L_{-2} \ket{0}$. }
\label{table:su2level5}
\end{table}
We have a relation of the form 
\begin{align}
 L_{-5} \sim J_{-3}^0 J_{-2}^0 + J_{-3}^+ J_{-2}^- + J_{-3}^- J_{-2}^+ +J_{-4}^0 J_{-1}^0  + J_{-4}^+ J_{-1}^- + J_{-4}^- J_{-1}^+ +  \ldots \ . 
\end{align}
It is straightforward (albeit tedious) to compute the refined character following the same procedure as before.  

Explicit computation of null states become rather cumbersome at this level. But we can compute the refined character assuming there is no null state, which is the case for the $(A_1, D_{2n+1})$ theory with $n \ge 3$. If there is no null state, then we get the refined character at this level as
\begin{align}
\begin{split}
Z^{(5)}=  q^5 \big( T(\chi_1 +\chi_3) +T^2 (2\chi_1 + 5 \chi_3 + 2\chi_5) + T^3(3\chi_3 +4 \chi_5 + 2 \chi_7) + T^4 (2\chi_7 + \chi_9) + T^5 \chi_{11} \big), 
\end{split}
\end{align}
which agrees with the Macdonald index of $(A_1, D_{2n+1})$ theory for $n \ge 3$. 

\paragraph{A closed form formula for the case without null state}
So far we have computed the refined character in a brute force way. If there is no null state in the vacuum module, it is possible to write a simple closed form formula for the refined character. To this end, note that we have four generators $J^+_{-n}, J^0_{-n}, J^-_{-n}, L_{-n}$ before imposing the relations among $L$ and $J$'s. Each of them contributes to the character by $q^n T z^2, q^n T, q^n T z^{-2}, q^n T$ respectively. For the current generators $J^a_{-n}$, the modes with $n\ge1$ contribute. For the Virasoro generator $L_{-m}$, the modes with $m\ge2$ contribute since $L_{-1}\ket{0}$ has zero norm. Therefore, the vacuum module will be generated by various strings of operators formed out of $J^{\pm, 0}_{-n}$ and $L_{-m}$. On top of this, we have the Sugawara relation $L_{-n} \sim \sum_m J^a_{-n-m} J^a_m$ for each $n$. Therefore we should subtract the character by $q^n T^2$ for each $n\ge 2$ to compensate for the overcounting. 

One can use plethystic exponential to write the states created by the operators of `multiple letters' respecting the relations. We get
\begin{align} \label{eq:closedRef}
 Z_{\CV}^{\textrm{ref}} (z; q, T) = \PE \left[ \frac{q T}{1-q} \left( z^2 + \frac{1}{z^2} + 1 \right) + \frac{q^2 T}{1-q} - \frac{q^2 T^2}{1-q} \right]_{q, T, z} \ , 
\end{align}
where the plethystic exponential (PE) is defined as 
\begin{align}
 \PE [f(z_1, \cdots, z_n)]_{z_1, \cdots, z_n} = \exp \left[ \sum_{k = 1}^\infty \frac{f(z_1^k, \cdots, z_n^k)}{k} \right] \ .  
\end{align}
In the equation \eqref{eq:closedRef}, the first term inside the PE comes from the current generators. The second term comes from the Virasoro generators, and the last term comes from the Sugawara relation. This expression can be expanded to give
\begin{align}
 Z_{\CV}^{\textrm{ref}} (z; q, T) = \prod_{n=1}^{\infty} \frac{(1-q^{n+1} T^2)}{(1-q^n T z^2)(1-q^n T z^{-2})(1-q^n T )(1-q^{n+1} T)} \ , 
\end{align}
which agrees with the brute force computation done in this section. It would be interesting to incorporate the null states to show that the refined character is indeed given by the Macdonald index formulas as given in appendix \ref{sec:MacD}. 


\section{Lagrangian theories} \label{sec:Lag}

In this section, we discuss the superconformal theories constructed from gauge theory. Let us first start with the free theories. 

\paragraph{Free hypermultiplet from symplectic boson and $\asu(2)_{-\half}$}

The Macdonald index for a free hypermultiplet is given by
\begin{align} \label{eq:IdxHyper}
\begin{split}
 I_{\textrm{hyp}} &= \frac{1}{(t^\half z; q)(t^\half z^{-1}; q)} \\
 &= 1 + q T \chi_3  + q^2 \left(T (\chi_1 + \chi_3) + T^2 \chi_5 \right) + q^3 \left(T (2\chi_3 + \chi_1) + T^2 (\chi_5 + \chi_3) + T^3 \chi_7 \right)  \\
  &~~+  q^\half T^\half \chi_2 + q^{\frac{3}{2}} \left( T^\half \chi_2 + T^{\frac{3}{2}} \chi_4 \right) + q^{\frac{5}{2}} \left(T^\half \chi_2 + T^{\frac{3}{2}}(\chi_4 + \chi_2) + T^{\frac{5}{2}} \chi_6 \right) + \cdots \ , 
\end{split}
\end{align}
where we have separated the terms with integer powers in $q$ and half-integer powers in $q$. The chiral algebra for the free hypermultiplet is given by symplectic bosons (or equivalently complex free fermions with bosonic statistics). There are two operators $q(z), \tilde{q}(z)$ with OPE
\begin{align}
 q(z) \tilde{q}(0) = - \tilde{q}(z) q(0) \sim \frac{1}{z}  \ . 
\end{align}
One can expand the holomorphic fields $(q(z), \tilde{q}(z))$ in Laurent series 
\begin{align}
 q(z) = \sum_{n \in \IZ} \frac{q_n}{z^{n+\half}} \ , \qquad  \tilde{q}(z) = \sum_{n \in \IZ} \frac{\tilde{q}_n}{z^{n+\half}} \ . 
\end{align}
In terms of these modes, the descendants states of the vacuum module can be obtained by acting products of the negative modes on the vacuum. 
Now, the refined character for this chiral algebra can be easily computed to give
\begin{align}
 Z_{s.b} (q, T) = \frac{1}{(q^\half T^\half z; q)(q^\half T^\half z^{-1}; q)} = \frac{1}{(t^\half z^\pm; q)} \ , 
\end{align}
where we introduced the fugacity $z$ for the $U(1)$ rotating the $q$ and $\tilde{q}$ with opposite charges. We set $w(q)=w(\tilde{q})=\half$, which is the same as the scaling dimension of the symplectic bosons. The fugacity $T$ counts the (half of the) number of $q_n$ or $\tilde{q}_n$'s. This is exactly the same as the Macdonald index of a hypermultiplet. 

Let us make a remark that even though the free hypermultiplet theory has $\asu(2)_{-\half}$ symmetry, it does not mean the chiral algebra (or vertex operator algebra) is simply given by the vacuum module of the affine Lie algebra $\asu(2)_{-\half}$. 
The vacuum character for the $\asu(2)_{-\half}$ algebra is given by \cite{Kac:1988qc} (see also \cite{Ridout:2008nh})
\begin{align}
 \chi_{\hat{\CL}_0} (z; q) = \frac{1}{(q z^2; q)(z^{-2};q)(q; q)}\sum_{n \in \IZ} \left( z^{-6n} - z^{6n-2} \right) q^{2n(3n-1)} \ , 
\end{align}
where the first term in the denominator is coming from the characters of the Verma module generated by $J^{\pm, 0}_n$. 
In the Schur limit $t=q$, it turns out the index is identical to the sum of the characters of the vacuum module $\hat{\CL}_0$ and $j=\half$ module $\hat{\CL}_{\half}$, not just the vacuum character. This is the case because the chiral algebra given by the symplectic boson is larger than just the affine Kac-Moody algebra $\asu(2)_{-\half}$. But, it is worth mentioning that the vertex operator algebra for the symplectic boson is not even a direct sum of the vacuum and the $j=\half$ sector either. This can also be seen from the refined character for the $\widehat{\mathfrak{su}}(2)_{-\half}$ we computed in the previous section, which does not agree with the first line of \eqref{eq:IdxHyper}.

\paragraph{Free vector multiplet from $bc$-system}
The chiral algebra for a free vector is given by a $bc$-system with weights $(1, 0)$: 
\begin{align}
 b(z) c(0) = -c(z) b(0) \sim \frac{1}{z} 
\end{align}
The algebra does not include the zero mode of $c(z)$. They are generated by $b(z), \p c(z)$. We assign the fugacity $T$ to count the number of $b(z)$-modes, that is to set $w(b)=1$, $w(c)=0$. Then we get the partition function to be
\begin{align}
 Z_{bc} = (q; q)(q T; q) = (q; q)(t; q) \ , 
\end{align}
which is exactly the same as the Macdonald index of a vector multiplet. 

Up to this point the weight $w(X)$ for the generator $X$ seems to coincide with its scaling dimension. But this is not true in general, because the stress energy tensor is already an exception to this rule. The stress tensor $T(z)$ has dimension 2 but we assign $w(L)=1$. We will discuss the issue of the weight $w(X)$ more in section \ref{sec:conclusion}. 

\paragraph{Gauge theories}
The chiral algebra of the Lagrangian gauge theory at the zero coupling is simply given by the tensor product of the chiral algebras for the free vectors and hypermultiplets under the Gauss' law constraint. So the refined character can be obtained simply by taking the tensor product of that of the free theories and then integrating over the gauge group with Haar measure. This is exactly the same procedure to obtain the superconformal index for a Lagrangian theory. 

When the gauge coupling is non-zero, it was shown in \cite{Beem:2013sza} that the chiral algebra does not change since the Schur operator remains to be the Schur operator at finite coupling upon passing through the BRST cohomology. Therefore the refined character of the chiral algebra is not modified at finite coupling. 

Eventually, one should be able to find more transparent prescription for the refined character without referring to the 4d Lagrangian description. We leave this as a future work. 


\section{Conclusions and Open Problems} \label{sec:conclusion}
In this paper, we proposed a prescription to obtain the Macdonald index of a 4d $\CN=2$ SCFT from its associated 2d chiral algebra. In 2d language, we have come up with a notion of refined character for the vacuum module of the chiral algebra. We have performed a number of checks for the simplest examples. If our conjecture holds, it will give us a way to compute Macdonald index for various non-Lagrangian $\CN=2$ SCFTs. There are many questions remain to be answered. 

First of all, the dictionary is not complete without finding an inherently 2d way to extract the weight $w(X)$ of the fugacity $T$ for arbitrary generator $X$ in $\CA$. For the simplest case where the chiral algebra is simply Virasoro or the affine Kac-Moody with Sugawara stress tensor, there is essentially no issue of choosing the weight for a generator. But in general, there are more than one generators for the chiral algebra. For example, consider the $SU(2)$ class-$\CS$ theory given by the genus 2 UV curve without punctures. There are many generators of the same conformal dimension. and the $w(X)$ for such generators are not necessarily the same, as we can see from the table 6 of \cite{Beem:2013sza}. If we know the 4d theory, the weight is determined by $w(X)=R-r$ as we can see from the definition of the Macdonald index \eqref{eq:MacIdxDef}. But if there is a genuine 2d meaning to this `refinement,' there should be an independent method to deduce $w(X)$ without referring to the 4d origin of the chiral algebra. 

If our prescription is indeed valid, then there must be a way to understand why such a rule works. We have only provided some evidence that a refinement for the character is possible, but have not provided physics reason behind this relation. This would be one of the most important questions that needs to be addressed. 

The notion of the refined character is not restricted to the vacuum modules. One can consider non-vacuum modules as well. It has been demonstrated that the Schur index with a line operator insertion reproduces a combination of other characters of the chiral algebra \cite{Cordova:2016uwk}. 
Also, inserting the surface defect at a point of the plane of the chiral algebra produces other characters of the chiral algebra \cite{Cordova:2017mhb, BPR2017}. Another interesting phenomenon is that vacuum and non-vacuum characters appear as a solution to the modular differential equation corresponding to the chiral algebra \cite{Beem:2017ooy, Arakawa:2016hkg}. It would be interesting to understand the effect of our refinement in this context as well. 

One of the consequences of our conjecture is that one can take a Hall-Littlewood limit of the refined character. It is to take $q\to 0$ while $t=qT$ fixed. For many theories\footnote{It fails for the (generalized) quiver gauge theories with loops.}, this limit of the index gives the Hilbert series of the Higgs branch. Therefore we obtain the space of holomorphic functions on the Higgs branch (or the Higgs branch chiral ring) as $\CV_H \equiv \bigoplus_{k\ge 0} V_{k}^{(k)}$. This space $\CV_H$ appears to be identical to the ring of holomorphic function $\IC[X_\CV]$ over the associated variety $X_\CV$ of the vertex algebra $\CV$ studied by Arakawa \cite{MR3456698, Arakawa:2016ac}. The connection between the Higgs branch and the associated variety was conjectured in \cite{Beem:2017ooy} and further tested in \cite{Song:2017oew}. It has been shown that every vertex algebra has a decreasing filtration, from which an associated graded vector space can be formed \cite{Li2005}. This is also related to the Zhu's $C_n$-algebra \cite{zhu1990vertex,zhu1996modular}. It would be interesting to clarify its connection to our grading. 

If we forget about the 4d origin, the refined character can be defined for any conformal field theories (if there is a way to determine the weight $w(X)$). One question to ask is if there is an interesting structure on its own. For example, does it transform nicely under the modular transformation? Can we also refine the modular differential equation?
Also, in the light of our original goal of understanding 4d $\CN=2$ theory, it would be interesting to see if it has any applications to the superconformal bootstrap program \cite{Beem:2014zpa,Lemos:2015awa}. 

There are various techniques available to compute ordinary character efficiently. In our case, we needed the explicit form of the null vectors to find the refined character. This can be done in a brute force way for our examples to low orders. We performed explicit computations for the first few orders. But it would be desirable to find a more efficient way to obtain the explicit form of the null vectors. This problem is essentially solved long time ago by Malikov-Feigin-Fuchs \cite{Malikov1986}, but their form is not directly applicable to our fractional level affine Kac-Moody algebra. This is because their expression is written in terms of fractional powers of generators of $\CA$, so one needs to convert this into the canonical form \cite{Bauer:1993jj,Mathieu:1998mb, Creutzig:2013yca}. It would be desirable to improve the brute force calculation done in the current paper. It may enable us to go beyond the simplest examples we studied here. 

Finally, we would like to point out a tantalizing connection between the BPS monodromy and the Macdonald index. As the Schur index (also with the insertion of defects) can be obtained from the BPS spectrum on the Coulomb branch \cite{Cordova:2015nma, Cecotti:2015lab, Cordova:2016uwk, Cordova:2017ohl, Cordova:2017mhb}, it may be possible to obtain the Macdonald index from the BPS monodromy.\footnote{One simple-minded `refinement' we can consider is to replace the BPS monodromy operator by 
$$
 \CM(q) =  T \left( \prod_{\g} \prod_{n = 0}^\infty (1-q^{n+s+\half} X_\g)^{(-1)^{2s}}\right) \to \CM(q, t) = T \left( \prod_{\g} \prod_{n = 0}^\infty (1-t^{s+\half} q^{n}  X_\g)^{(-1)^{2s}}\right) , 
$$
since it correctly reproduces the `Macdonald index' for the free hypermultiplets and $U(1)$ gauge theories. But we have not found a procedure that works for the interacting theories. }

\begin{acknowledgments}
The author would like to thank Christopher Beem, Sergio Cecotti, Martin Fluder, Ken Intriligator, John McGreevy, Leonardo Rastelli, Shu-Heng Shao, Cumrun Vafa, Dan Xie and Wenbin Yan for discussions and correspondence. He especially thanks Wenbin Yan for collaborating at the early stage of this work. He also thanks the anonymous referee who helped him to improve the presentation of the paper. He would like to thank Korea Institute for Advanced Study and Simons Center for Geometry and Physics during the 2016 Summer Workshop for hospitality. He would also like to thank the organizers of the workshop ``Exact Operator Algebras in Superconformal Field Theories" and Perimeter Institute for hospitality. 
This work is supported in part by the US Department of Energy under UCSD's contract de-sc0009919 and also by Hwa-Ahm foundation.
\end{acknowledgments}

\appendix

\section{Macdonald index of the Argyres-Douglas theory}
In this appendix, we review the Macdonald index of the AD theories that are studied in this paper. The superconformal index of an $\CN=2$ superconformal field theory is defined as
\begin{align}
 I (p, q, t; z) = \tr (-1)^F p^{j_1 - j_2 } q^{j_1 + j_2 } t^{R} \left( \frac{pq}{t} \right)^r \prod_{i} z_i^{F_i} \ ,  
\end{align}
where $j_1, j_2$ are the Cartans for the Lorentz group and $R$ and $r$ denotes $SU(2)_R$ and $U(1)_r$ generators. Here $F_i$ are the Cartans of the flavor symmetry. The trace is over $\frac{1}{8}$-BPS states satisfying $\delta_{1-} \equiv \Delta - 2j_1 - 2R - r = 0$. 
 
The Macdonald limit of the index is obtained by taking $p \to 0$. Then the trace is over $\frac{1}{4}$-BPS states, satisfying $j_1+j_2 - r = 0$ in addition to $\delta_{1-}=0$. The Schur limit is obtained by taking $p=0$ and $t=q$, but it gets contribution from the same set of states as the Macdonald index. The Hall-Littlewood limit (or the Higgs branch limit) is obtained by taking $p=q=0$. In this case, the trace is over $\frac{3}{8}$-BPS states satisfying $j_1=0$ in addition to the condition satisfied by the Schur operators. 

\subsection{$(A_1, A_{2n})$ theory} \label{sec:MacA}
Let us consider $(A_1, A_{2n})$ type Argyres-Douglas theory. The Macdonald index for this theory is conjectured to be given by \cite{Song:2015wta}
\begin{align} \label{eq:MacIdx}
 I_{(A_1, A_{2n})}(q, t) = \frac{1}{(t^2; q)} \sum_{\lambda \ge 0} (-1)^\lambda q^{\lambda(\lambda+1)(n+\frac{3}{2})} \left( \frac{t}{q} \right)^{\lambda(n+2)} \frac{(q^{\lambda+1}; q)_{\lambda} (t^2 q^{2 \lambda}; q)}{(t q^{\lambda}; q)_{\lambda} (tq^{2\lambda+1}; q) } [2\lambda]_{q, t}  \ , 
\end{align}
where 
\begin{align}
 [n]_{q, t} = \sum_{i=0}^{n} \frac{(t; q)_i (t; q)_{n-i}}{(q; q)_i (q; q)_{n-i}} t^{\frac{n}{2} - i} \ ,
\end{align}
and $(x; q)_n = \prod_{i=0}^{n-1} (1 - x q^i)$ and $(x; q) \equiv (x; q)_\infty$. 
This expression has been tested against the recent computation done using $\CN=1$ gauge theory realization \cite{Maruyoshi:2016tqk, Maruyoshi:2016aim}. 
Note that when $t=q$, $[n]_{q, t=q} = [n]_q \equiv \frac{q^{n/2} - q^{-n/2}}{q^{1/2}-q^{-1/2}}$. In the Schur limit $t=q$, the index simplifies to 
\begin{align}
 I_{(A_1, A_{2n})}(q) = \frac{1}{(q^2; q)} \sum_{\lambda \ge 0} (-1)^{\lambda} q^{\lambda(\lambda+1)(n+\frac{3}{2})} [2\lambda]_q \ . 
\end{align}
This expression is shown to be equal to the vacuum character of the Virasoro minimal model $\CM(2, 2n+3)$ \cite{Song:2017oew}, which can be written as \cite{MR1669957}
\begin{align}
 \chi_0^{(2, 2n+3)} = \frac{(q, q^{2n+2}, q^{2n+3}; q^{2n+3})}{(q; q)} = \prod_{i=1}^{2n} \frac{1}{(q^{i+1}; q^{2n+3})} \ , 
\end{align}
where $(x_1, x_2, \ldots, x_n; q) \equiv (x_1; q)(x_2;q) \ldots (x_n; q)$. 
Here we write explicit expressions of the Macdonald indices of $(A_1, A_{2n})$ theory for $n=1, 2, 3$: 
\begin{align}
\begin{split}
 I_{(A_1, A_2)} &= 1 + q^2 T + q^3 T + q^4 T + q^5 T + q^6 (T + T^2) + q^7 (T + T^2) \\
 & \qquad \qquad \qquad \qquad + q^8 (2T^2 + T) + q^9 (T+2T^2) + q^{10} (T+3T^2)  + \ldots ,  
\end{split} \\ \nn \\
\begin{split}
 I_{(A_1, A_4)} &= 1 + q^2 T + q^3 T + q^4 (T+T^2) + q^5 (T+T^2) + q^6 (T+2T^2) + q^7 (T + 2 T^2) \\
 & \qquad   + q^8 (T+3T^2 +T^3) + q^9 (T + 3T^2 + 2 T^3) + q^{10} (T+4T^2 +3T^3) + \ldots ,  
\end{split} \\ \nn \\
\begin{split}
 I_{(A_1, A_6)} &= 1 + q^2 T + q^3 T + q^4 (T+T^2) + q^5 (T+T^2) + q^6 (T+2T^2+ T^3)  \\ 
  & \qquad  + q^7 (T + 2 T^2 + T^3) + (T+3T^2 + 2T^3) q^8 + (1 +3T+3T^2) q^9 +\ldots \ . 
\end{split}
\end{align}
Here we used $t=qT$. These expression reduces to the vacuum character of the $\CM(2, 2n+3)$ model upon taking $T \to 1$. 

\subsection{$(A_1, D_{2n+1})$ theory} \label{sec:MacD}
The Macdonald index for the AD theory of type $(A_1, D_{2n+1})$ is conjectured to be given by
\begin{align}
 I_{(A_1, D_{2n+1})} = \frac{1}{(t z^{\pm 2, 0})_\infty} \sum_{\lambda \ge 0} (-1)^\lambda q^{\lambda(\lambda+1)(n+\frac{1}{2})} \left(\frac{t}{q}\right)^{\lambda(n+1)} \frac{(q^{\lambda+1})_{\lambda} (t^2 q^{2 \lambda})_\infty}{(t q^{\lambda})_{\lambda} (tq^{2\lambda+1})_\infty } P_{2\lambda}(z) ,
\end{align}
where we used abbreviation $(x)_n \equiv (x; q)_n$. The $SU(2)$ Macdonald polynomial is given by
\begin{align}
 P_{\lambda}(z) = \sum_{i=0}^\lambda \frac{(t; q)_i}{(q; q)_i} \frac{(t; q)_{\lambda-i}}{(q; q)_{\lambda-i}} z^{2i - \lambda} \ . 
\end{align}
The chiral algebra of $(A_1, D_{2n+1})$ is expected to be $\widehat{\mathfrak{su}}(2)_{-\frac{4n}{2n+1}}$. This expression agrees with the expression obtained from the $\CN=1$ gauge theory flowing to the AD theory \cite{Agarwal:2016pjo}. 

In the Schur limit $t \to q$, the index can be written concisely in terms of Plethystic Exponential (PE) or a product formula \cite{Song:2017oew}
\begin{align}
 I_{(A_1, D_{2n+1})} = \PE \left[ \frac{q-q^{2n+1}}{(1-q)(1-q^{2n+1})} \chi_{\textrm{adj}(z)} \right]_{q, z}
  =  \prod_{m \ge 0} \prod_{\a \in \{-2, 0, 2\}} \frac{(q^{(m-1)(2n+1)} z^{\a}; q)}{(q^{m(2n+1)+1} z^{\a}; q)} \ , 
\end{align}
where $\PE[a z]_z = (\frac{1}{1-z})^a$ and $\chi_{\textrm{adj}}(z) = z^2 +\frac{1}{z^2}+1$ is the character of the adjoint representation of the flavor symmetry $\mathfrak{su}(2)$.  This gives the vacuum character of the corresponding chiral algebra. 

Here we write explicit expressions of the Macdonald indices of $(A_1, D_{2n+1})$ theory for $n=1, 2, 3$: 
\begin{align}
\begin{split} \label{eq:MacA1D3}
 I_{(A_1, D_3)} =& ~1 + q T \chi_3 + q^2 \left( T (\chi_3 + \chi_1) + T^2 \chi_5 \right) \\
 &{ } + q^3 \left( T (\chi_3 + \chi_1) + T^2 (\chi_5 + \boxed{\chi_3}) + T^3 \chi_7 \right) \\ 
 &{ } + q^4 \left( T(\chi_3 +\chi_1) +T^2 (2 \chi_5 + 2\chi_3 + \chi_1) + T^3(\chi_7 + \chi_5 ) + T^4 \chi_9 \right) \\
 &{ } + q^5 \big( T(\chi_3 + \chi_1) + T^2 (2\chi_5 + 4\chi_3 + \chi_1) + T^3(2\chi_7 + 2 \chi_5 + \chi_3) \\ 
 &{ }  \qquad \qquad + T^4(\chi_9 + \chi_7) + T^5\chi_{11} \big) + O(q^6) \ , 
\end{split} 
\end{align}
\begin{align}
\begin{split}\label{eq:MacA1D5}
 I_{(A_1, D_5)}
  =& ~1 + q T \chi_3 + q^2 \left(T (\chi_3  + \chi_1 ) + T^2 \chi_5 \right)  \\
  &{ } + q^3 \left( T(\chi_3 + \chi_1) + T^2 (\chi_5 + 2 \chi_3) + T^3 \chi_7 \right) \\ 
  &{ } + q^4 \left( T(\chi_3 + \chi_1) + T^2 (2 \chi_5 + 3 \chi_3 + 2 \chi_1) + T^3 (\chi_7 + 2 \chi_5) + T^4 \chi_9 \right) \\
  &{ } + q^5 \big( T(\chi_3 + \chi_1) + T^2 (2\chi_5 + 5 \chi_3 + 2 \chi_1) + T^3 (2 \chi_7 + 4\chi_5 + \boxed{2 \chi_3})  \\
  &{ } \qquad \qquad + T^4 (\chi_9 + 2 \chi_7) + T^5 \chi_{11} \big) + O(q^6) \ ,  
\end{split} 
\end{align}
\begin{align}
\begin{split}  \label{eq:MacA1D7}
 I_{(A_1, D_7)}
  =& ~1 + q T \chi_3 + q^2 \left(T (\chi_3  + \chi_1 ) + T^2 \chi_5 \right)  \\
  &{ } + q^3 \left( T(\chi_3 + \chi_1) + T^2 (\chi_5 + 2 \chi_3) + T^3 \chi_7 \right) \\ 
  &{ } + q^4 \left( T(\chi_3 + \chi_1) + T^2 (2 \chi_5 + 3 \chi_3 + 2 \chi_1) + T^3 (\chi_7 + 2 \chi_5) + T^4 \chi_9 \right) \\
  &{ } + q^5 \big( T(\chi_3 + \chi_1) + T^2 (2\chi_5 + 5 \chi_3 + 2 \chi_1) + T^3 (2\chi_7 + 4 \chi_5 + 3 \chi_3) \\
  & { } \qquad \qquad + T^4(\chi_9 + 2\chi_7) + T^5 \chi_{11}  \big) + O(q^6) \ ,  
\end{split}
\end{align}
which should be reduced to the vacuum character of $\widehat{\mathfrak{su}}(2)_{-\frac{4}{3}}, \widehat{\mathfrak{su}}(2)_{-\frac{8}{5}}, \widehat{\mathfrak{su}}(2)_{-\frac{12}{7}}$ upon taking $T \to 1$ respectively. The boxed terms denote the first term that starts to differ from the $(A_1, D_{2n+1})$ with higher $n$. The coefficient of the boxed term signals existence of a null vector in the corresponding vertex algebra. 


\bibliographystyle{jhep}
\bibliography{refs}

\providecommand{\href}[2]{#2}\begingroup\raggedright\begin{thebibliography}{10}

\bibitem{Beem:2013sza}
C.~Beem, M.~Lemos, P.~Liendo, W.~Peelaers, L.~Rastelli and B.~C. van Rees,
  \emph{{Infinite Chiral Symmetry in Four Dimensions}},
  \href{https://doi.org/10.1007/s00220-014-2272-x}{\emph{Commun. Math. Phys.}
  {\bfseries 336} (2015) 1359--1433},
  [\href{https://arxiv.org/abs/1312.5344}{{\ttfamily 1312.5344}}].

\bibitem{Beem:2014rza}
C.~Beem, W.~Peelaers, L.~Rastelli and B.~C. van Rees, \emph{{Chiral algebras of
  class S}}, \href{https://doi.org/10.1007/JHEP05(2015)020}{\emph{JHEP}
  {\bfseries 05} (2015) 020},
  [\href{https://arxiv.org/abs/1408.6522}{{\ttfamily 1408.6522}}].

\bibitem{Lemos:2014lua}
M.~Lemos and W.~Peelaers, \emph{{Chiral Algebras for Trinion Theories}},
  \href{https://doi.org/10.1007/JHEP02(2015)113}{\emph{JHEP} {\bfseries 02}
  (2015) 113}, [\href{https://arxiv.org/abs/1411.3252}{{\ttfamily 1411.3252}}].

\bibitem{Cordova:2015nma}
C.~Cordova and S.-H. Shao, \emph{{Schur Indices, BPS Particles, and
  Argyres-Douglas Theories}},
  \href{https://doi.org/10.1007/JHEP01(2016)040}{\emph{JHEP} {\bfseries 01}
  (2016) 040}, [\href{https://arxiv.org/abs/1506.00265}{{\ttfamily
  1506.00265}}].

\bibitem{Buican:2015ina}
M.~Buican and T.~Nishinaka, \emph{{On the superconformal index of
  Argyres--Douglas theories}},
  \href{https://doi.org/10.1088/1751-8113/49/1/015401}{\emph{J. Phys.}
  {\bfseries A49} (2016) 015401},
  [\href{https://arxiv.org/abs/1505.05884}{{\ttfamily 1505.05884}}].

\bibitem{Xie:2016evu}
D.~Xie, W.~Yan and S.-T. Yau, \emph{{Chiral algebra of Argyres-Douglas theory
  from M5 brane}},  \href{https://arxiv.org/abs/1604.02155}{{\ttfamily
  1604.02155}}.

\bibitem{Nishinaka:2016hbw}
T.~Nishinaka and Y.~Tachikawa, \emph{{On 4d rank-one N=3 superconformal field
  theories}},  \href{https://arxiv.org/abs/1602.01503}{{\ttfamily 1602.01503}}.

\bibitem{Lemos:2016xke}
M.~Lemos, P.~Liendo, C.~Meneghelli and V.~Mitev, \emph{{Bootstrapping
  $\mathcal{N}=3$ superconformal theories}},
  \href{https://arxiv.org/abs/1612.01536}{{\ttfamily 1612.01536}}.

\bibitem{Bonetti:2016nma}
F.~Bonetti and L.~Rastelli, \emph{{Supersymmetric Localization in AdS$_5$ and
  the Protected Chiral Algebra}},
  \href{https://arxiv.org/abs/1612.06514}{{\ttfamily 1612.06514}}.

\bibitem{Liendo:2015ofa}
P.~Liendo, I.~Ramirez and J.~Seo, \emph{{Stress-tensor OPE in $ \mathcal{N}=2 $
  superconformal theories}},
  \href{https://doi.org/10.1007/JHEP02(2016)019}{\emph{JHEP} {\bfseries 02}
  (2016) 019}, [\href{https://arxiv.org/abs/1509.00033}{{\ttfamily
  1509.00033}}].

\bibitem{Lemos:2015orc}
M.~Lemos and P.~Liendo, \emph{{$\mathcal{N}=2$ central charge bounds from $2d$
  chiral algebras}}, \href{https://doi.org/10.1007/JHEP04(2016)004}{\emph{JHEP}
  {\bfseries 04} (2016) 004},
  [\href{https://arxiv.org/abs/1511.07449}{{\ttfamily 1511.07449}}].

\bibitem{Buican:2016arp}
M.~Buican and T.~Nishinaka, \emph{{Conformal Manifolds in Four Dimensions and
  Chiral Algebras}},  \href{https://arxiv.org/abs/1603.00887}{{\ttfamily
  1603.00887}}.

\bibitem{Beem:2014kka}
C.~Beem, L.~Rastelli and B.~C. van Rees, \emph{{$ \mathcal{W} $ symmetry in six
  dimensions}}, \href{https://doi.org/10.1007/JHEP05(2015)017}{\emph{JHEP}
  {\bfseries 05} (2015) 017},
  [\href{https://arxiv.org/abs/1404.1079}{{\ttfamily 1404.1079}}].

\bibitem{Kinney:2005ej}
J.~Kinney, J.~M. Maldacena, S.~Minwalla and S.~Raju, \emph{{An Index for 4
  dimensional super conformal theories}},
  \href{https://doi.org/10.1007/s00220-007-0258-7}{\emph{Commun. Math. Phys.}
  {\bfseries 275} (2007) 209--254},
  [\href{https://arxiv.org/abs/hep-th/0510251}{{\ttfamily hep-th/0510251}}].

\bibitem{Romelsberger:2005eg}
C.~Romelsberger, \emph{{Counting chiral primaries in N = 1, d=4 superconformal
  field theories}},
  \href{https://doi.org/10.1016/j.nuclphysb.2006.03.037}{\emph{Nucl. Phys.}
  {\bfseries B747} (2006) 329--353},
  [\href{https://arxiv.org/abs/hep-th/0510060}{{\ttfamily hep-th/0510060}}].

\bibitem{Gadde:2011ik}
A.~Gadde, L.~Rastelli, S.~S. Razamat and W.~Yan, \emph{{The 4d Superconformal
  Index from q-deformed 2d Yang-Mills}},
  \href{https://doi.org/10.1103/PhysRevLett.106.241602}{\emph{Phys.Rev.Lett.}
  {\bfseries 106} (2011) 241602},
  [\href{https://arxiv.org/abs/1104.3850}{{\ttfamily 1104.3850}}].

\bibitem{Gadde:2011uv}
A.~Gadde, L.~Rastelli, S.~S. Razamat and W.~Yan, \emph{{Gauge Theories and
  Macdonald Polynomials}},
  \href{https://doi.org/10.1007/s00220-012-1607-8}{\emph{Commun.Math.Phys.}
  {\bfseries 319} (2013) 147--193},
  [\href{https://arxiv.org/abs/1110.3740}{{\ttfamily 1110.3740}}].

\bibitem{Argyres:1995jj}
P.~C. Argyres and M.~R. Douglas, \emph{{New phenomena in SU(3) supersymmetric
  gauge theory}},
  \href{https://doi.org/10.1016/0550-3213(95)00281-V}{\emph{Nucl. Phys.}
  {\bfseries B448} (1995) 93--126},
  [\href{https://arxiv.org/abs/hep-th/9505062}{{\ttfamily hep-th/9505062}}].

\bibitem{Argyres:1995xn}
P.~C. Argyres, M.~R. Plesser, N.~Seiberg and E.~Witten, \emph{{New N=2
  superconformal field theories in four-dimensions}},
  \href{https://doi.org/10.1016/0550-3213(95)00671-0}{\emph{Nucl. Phys.}
  {\bfseries B461} (1996) 71--84},
  [\href{https://arxiv.org/abs/hep-th/9511154}{{\ttfamily hep-th/9511154}}].

\bibitem{feigin2009pbw}
E.~Feigin, \emph{{The PBW filtration}}, {\emph{Representation Theory of the
  American Mathematical Society} {\bfseries 13} (2009) 165--181}.

\bibitem{Xie:2012hs}
D.~Xie, \emph{{General Argyres-Douglas Theory}},
  \href{https://doi.org/10.1007/JHEP01(2013)100}{\emph{JHEP} {\bfseries 1301}
  (2013) 100}, [\href{https://arxiv.org/abs/1204.2270}{{\ttfamily 1204.2270}}].

\bibitem{Xie:2013jc}
D.~Xie and P.~Zhao, \emph{{Central charges and RG flow of strongly-coupled N=2
  theory}}, \href{https://doi.org/10.1007/JHEP03(2013)006}{\emph{JHEP}
  {\bfseries 03} (2013) 006},
  [\href{https://arxiv.org/abs/1301.0210}{{\ttfamily 1301.0210}}].

\bibitem{Wang:2015mra}
Y.~Wang and D.~Xie, \emph{{Classification of Argyres-Douglas theories from M5
  branes}}, \href{https://doi.org/10.1103/PhysRevD.94.065012}{\emph{Phys. Rev.}
  {\bfseries D94} (2016) 065012},
  [\href{https://arxiv.org/abs/1509.00847}{{\ttfamily 1509.00847}}].

\bibitem{Buican:2015tda}
M.~Buican and T.~Nishinaka, \emph{{Argyres-Douglas Theories, the Macdonald
  Index, and an RG Inequality}},
  \href{https://doi.org/10.1007/JHEP02(2016)159}{\emph{JHEP} {\bfseries 02}
  (2016) 159}, [\href{https://arxiv.org/abs/1509.05402}{{\ttfamily
  1509.05402}}].

\bibitem{Song:2015wta}
J.~Song, \emph{{Superconformal indices of generalized Argyres-Douglas theories
  from 2d TQFT}}, \href{https://doi.org/10.1007/JHEP02(2016)045}{\emph{JHEP}
  {\bfseries 02} (2016) 045},
  [\href{https://arxiv.org/abs/1509.06730}{{\ttfamily 1509.06730}}].

\bibitem{Song:2017oew}
J.~Song, D.~Xie and W.~Yan, \emph{{Vertex operator algebras of Argyres-Douglas
  theories from M5-branes}},
  \href{https://arxiv.org/abs/1706.01607}{{\ttfamily 1706.01607}}.

\bibitem{Cecotti:2015lab}
S.~Cecotti, J.~Song, C.~Vafa and W.~Yan, \emph{{Superconformal Index, BPS
  Monodromy and Chiral Algebras}},
  \href{https://arxiv.org/abs/1511.01516}{{\ttfamily 1511.01516}}.

\bibitem{Cordova:2016uwk}
C.~Cordova, D.~Gaiotto and S.-H. Shao, \emph{{Infrared Computations of Defect
  Schur Indices}}, \href{https://doi.org/10.1007/JHEP11(2016)106}{\emph{JHEP}
  {\bfseries 11} (2016) 106},
  [\href{https://arxiv.org/abs/1606.08429}{{\ttfamily 1606.08429}}].

\bibitem{Cecotti:2010fi}
S.~Cecotti, A.~Neitzke and C.~Vafa, \emph{{R-Twisting and 4d/2d
  Correspondences}},  \href{https://arxiv.org/abs/1006.3435}{{\ttfamily
  1006.3435}}.

\bibitem{Iqbal:2012xm}
A.~Iqbal and C.~Vafa, \emph{{BPS Degeneracies and Superconformal Index in
  Diverse Dimensions}},
  \href{https://doi.org/10.1103/PhysRevD.90.105031}{\emph{Phys. Rev.}
  {\bfseries D90} (2014) 105031},
  [\href{https://arxiv.org/abs/1210.3605}{{\ttfamily 1210.3605}}].

\bibitem{Maruyoshi:2016tqk}
K.~Maruyoshi and J.~Song, \emph{{Enhancement of Supersymmetry via
  Renormalization Group Flow and the Superconformal Index}},
  \href{https://doi.org/10.1103/PhysRevLett.118.151602}{\emph{Phys. Rev. Lett.}
  {\bfseries 118} (2017) 151602},
  [\href{https://arxiv.org/abs/1606.05632}{{\ttfamily 1606.05632}}].

\bibitem{Maruyoshi:2016aim}
K.~Maruyoshi and J.~Song, \emph{{$ \mathcal{N}=1 $ deformations and RG flows of
  $ \mathcal{N}=2 $ SCFTs}},
  \href{https://doi.org/10.1007/JHEP02(2017)075}{\emph{JHEP} {\bfseries 02}
  (2017) 075}, [\href{https://arxiv.org/abs/1607.04281}{{\ttfamily
  1607.04281}}].

\bibitem{Agarwal:2016pjo}
P.~Agarwal, K.~Maruyoshi and J.~Song, \emph{{$ \mathcal{N} $ =1 Deformations
  and RG flows of $ \mathcal{N} $ =2 SCFTs, part II: non-principal
  deformations}}, \href{https://doi.org/10.1007/JHEP12(2016)103,
  10.1007/JHEP04(2017)113}{\emph{JHEP} {\bfseries 12} (2016) 103},
  [\href{https://arxiv.org/abs/1610.05311}{{\ttfamily 1610.05311}}].

\bibitem{Gaberdiel:2001ny}
M.~R. Gaberdiel, \emph{{Fusion rules and logarithmic representations of a WZW
  model at fractional level}},
  \href{https://doi.org/10.1016/S0550-3213(01)00490-4}{\emph{Nucl. Phys.}
  {\bfseries B618} (2001) 407--436},
  [\href{https://arxiv.org/abs/hep-th/0105046}{{\ttfamily hep-th/0105046}}].

\bibitem{Kac:1988qc}
V.~G. Kac and M.~Wakimoto, \emph{{Modular invariant representations of infinite
  dimensional Lie algebras and superalgebras}},
  \href{https://doi.org/10.1073/pnas.85.14.4956}{\emph{Proc. Nat. Acad. Sci.}
  {\bfseries 85} (1988) 4956--5960}.

\bibitem{Ridout:2008nh}
D.~Ridout, \emph{{$\widehat{sl}(2)_{-1/2}$: A Case Study}},
  \href{https://doi.org/10.1016/j.nuclphysb.2009.01.008}{\emph{Nucl. Phys.}
  {\bfseries B814} (2009) 485--521},
  [\href{https://arxiv.org/abs/0810.3532}{{\ttfamily 0810.3532}}].

\bibitem{Cordova:2017mhb}
C.~Cordova, D.~Gaiotto and S.-H. Shao, \emph{{Surface Defects and Chiral
  Algebras}}, \href{https://doi.org/10.1007/JHEP05(2017)140}{\emph{JHEP}
  {\bfseries 05} (2017) 140},
  [\href{https://arxiv.org/abs/1704.01955}{{\ttfamily 1704.01955}}].

\bibitem{BPR2017}
C.~Beem, W.~Peelaers and L.~Rastelli.

\bibitem{Beem:2017ooy}
C.~Beem and L.~Rastelli, \emph{{Vertex operator algebras, Higgs branches, and
  modular differential equations}},
  \href{https://arxiv.org/abs/1707.07679}{{\ttfamily 1707.07679}}.

\bibitem{Arakawa:2016hkg}
T.~Arakawa and K.~Kawasetsu, \emph{{Quasi-lisse vertex algebras and modular
  linear differential equations}},
  \href{https://arxiv.org/abs/1610.05865}{{\ttfamily 1610.05865}}.

\bibitem{MR3456698}
T.~Arakawa, \emph{Associated varieties of modules over {K}ac-{M}oody algebras
  and {$C_2$}-cofiniteness of {$W$}-algebras}, {\emph{Int. Math. Res. Not.
  IMRN} (2015) 11605--11666}.

\bibitem{Arakawa:2016ac}
T.~Arakawa, \emph{Introduction to w-algebras and their representation theory},
  \href{https://arxiv.org/abs/1605.00138}{{\ttfamily 1605.00138}}.

\bibitem{Li2005}
H.~Li, \emph{Abelianizing vertex algebras},
  \href{https://doi.org/10.1007/s00220-005-1348-z}{\emph{Communications in
  Mathematical Physics} {\bfseries 259} (2005) 391--411}.

\bibitem{zhu1990vertex}
Y.~Zhu, \emph{Vertex Operator Algebras, Elliptic Functions and Modular Forms,
  Dissertation}.
\newblock Yale University, 1990.

\bibitem{zhu1996modular}
Y.~Zhu, \emph{Modular invariance of characters of vertex operator algebras},
  {\emph{Journal of the American Mathematical Society} {\bfseries 9} (1996)
  237--302}.

\bibitem{Beem:2014zpa}
C.~Beem, M.~Lemos, P.~Liendo, L.~Rastelli and B.~C. van Rees, \emph{{The $
  \mathcal{N}=2 $ superconformal bootstrap}},
  \href{https://doi.org/10.1007/JHEP03(2016)183}{\emph{JHEP} {\bfseries 03}
  (2016) 183}, [\href{https://arxiv.org/abs/1412.7541}{{\ttfamily 1412.7541}}].

\bibitem{Lemos:2015awa}
M.~Lemos and P.~Liendo, \emph{{Bootstrapping $ \mathcal{N}=2 $ chiral
  correlators}}, \href{https://doi.org/10.1007/JHEP01(2016)025}{\emph{JHEP}
  {\bfseries 01} (2016) 025},
  [\href{https://arxiv.org/abs/1510.03866}{{\ttfamily 1510.03866}}].

\bibitem{Malikov1986}
F.~G. Malikov, B.~L. Feigin and D.~B. Fuks, \emph{Singular vectors in verma
  modules over kac-moody algebras},
  \href{https://doi.org/10.1007/BF01077264}{\emph{Functional Analysis and Its
  Applications} {\bfseries 20} (1986) 103--113}.

\bibitem{Bauer:1993jj}
M.~Bauer and N.~Sochen, \emph{{Fusion and singular vectors in A1(1) highest
  weight cyclic modules}},
  \href{https://doi.org/10.1007/BF02097060}{\emph{Commun. Math. Phys.}
  {\bfseries 152} (1993) 127--160},
  [\href{https://arxiv.org/abs/hep-th/9201079}{{\ttfamily hep-th/9201079}}].

\bibitem{Mathieu:1998mb}
P.~Mathieu and M.~A. Walton, \emph{{On principal admissible representations and
  conformal field theory}},
  \href{https://doi.org/10.1016/S0550-3213(99)00252-7}{\emph{Nucl. Phys.}
  {\bfseries B553} (1999) 533--558},
  [\href{https://arxiv.org/abs/hep-th/9812192}{{\ttfamily hep-th/9812192}}].

\bibitem{Creutzig:2013yca}
T.~Creutzig and D.~Ridout, \emph{{Modular Data and Verlinde Formulae for
  Fractional Level WZW Models II}},
  \href{https://doi.org/10.1016/j.nuclphysb.2013.07.008}{\emph{Nucl. Phys.}
  {\bfseries B875} (2013) 423--458},
  [\href{https://arxiv.org/abs/1306.4388}{{\ttfamily 1306.4388}}].

\bibitem{Cordova:2017ohl}
C.~Cordova, D.~Gaiotto and S.-H. Shao, \emph{{Surface Defect Indices and 2d-4d
  BPS States}},  \href{https://arxiv.org/abs/1703.02525}{{\ttfamily
  1703.02525}}.

\bibitem{MR1669957}
G.~E. Andrews, A.~Schilling and S.~O. Warnaar, \emph{An {$A_2$} {B}ailey lemma
  and {R}ogers-{R}amanujan-type identities},
  \href{https://doi.org/10.1090/S0894-0347-99-00297-0}{\emph{J. Amer. Math.
  Soc.} {\bfseries 12} (1999) 677--702}.

\end{thebibliography}\endgroup

\end{document}